\documentclass[12pt]{article}
\usepackage{psfig}
\begin{document}
\title{\bf Can a non-ideal metal ferromagnet 
inject spin into a semiconductor with 100\% efficiency without a 
tunnel barrier?}
\author{
J. Wan, M. Cahay 
\\Department of Electrical and Computer Engineering and Computer Science\\
University of Cincinnati, Cincinnati, Ohio 45221\\
\\S. Bandyopadhyay \\
Department of Electrical and Computer Engineering \\
Virginia Commonwealth University, 
Richmond, Virginia 23284}
\date{}

\maketitle

\baselineskip=24pt

\begin{center}
{\bf ABSTRACT}
\end{center}

\bigskip

Current understanding of spin injection tells us that a metal 
ferromagnet can inject spin into a semiconductor with 100\% efficiency if 
either the ferromagnet is an ideal half metal with 100\% spin polarization,
or there exists a suitable tunnel barrier at the interface.
In this paper, we show that, 
at absolute zero temperature, 100\% spin injection efficiency 
from a {\it non-ideal} metal ferromagnet into a semiconductor quantum wire
can be reached at certain injection energies, {\it without} a tunnel barrier,
provided there is an axial magnetic field along the direction of current flow as well as a
spin orbit interaction in the semiconductor. 
At these injection energies, spin is injected {\it only} from the majority spin band 
of the ferromagnetic contact, resulting in 100\% spin injection efficiency.
This happens because of the presence of antiresonances in the transmission coefficient 
of the minority spins when their incident energies coincide with Zeeman energy 
states in the quantum wire.
At absolute zero and below a critical value of the axial magnetic field,
there are two distinct Zeeman energy states and therefore two injection energies 
at which ideal spin filtering is possible; above the
critical magnetic field there is only one such injection energy. The spin injection efficiency
rapidly decreases as the temperature increases. 
The rate of decrease is slower when the magnetic field
is above the critical value. 
The appropriate choice of semiconductor materials and structures necessary to maintain a 
large spin injection efficiency
at elevated temperatures is discussed.
  
%
%

\newpage
\vskip .2in
\begin{center}
{\bf I. INTRODUCTION} 
\end{center}

The problem of spin injection across a ferromagnetic/semiconductor (Fe/Sm) interface has received 
increasing attention over the last ten years with the advent of spintronics.
Several recent experimental investigations have shown successful
spin injection into semiconductor heterostructures from ferromagnetic metals
using tunnel barriers in the form of Schottky contacts \cite{zhu,hanbicki1,hanbicki2}, 
thin metal oxides \cite{motsnyi,manago,harris}, 
or AlAs barriers \cite{chun}. A spin injection efficiency of about 70 $\% $ has been
recently demonstrated using a CoFe/MgO tunnel contact to a GaAs layer \cite{harris}.
Simultaneously, several theoretical models have been
developed to understand the mechanisms controlling spin injection across specular and disordered 
Fe/Sm interfaces, some of which have been based on a simple Stoner model 
of the ferromagnetic contact \cite{cmhu,rashba,molenkamp}
while others have included the full electronic band structure of the contact \cite{zwierzycki,wunnicke}.

It is currently believed that if the ferromagnet is metallic and there 
is no tunnel barrier between the ferromagnet
and the semiconductor, then the infamous ``resistance mismatch'' problem
will preclude a high spin injection efficiency unless the ferromagnet is an
ideal half metal with 100 $\%$ spin polarization \cite{rashba,schmid}. 
Here we show that 100 $\%$ spin injection efficiency is possible at absolute zero 
temperature from a metallic ferromagnet with less than 100
$\%$ spin polarization into a semiconductor quantum wire, 
in spite of a resistance mismatch and in spite of the 
absence of any tunnel barrier, as long as there is an axial 
magnetic field along the wire and a Rashba spin orbit interaction \cite{rashba1}
in the semiconductor. The 100\% efficiency is obtained only for certain 
injection energies. At temperatures $T$ $>$ 0 K,  thermal smearing
will cause the energy-averaged spin injection efficiency to 
be much less than 100\%. This problem can be mitigated by
injecting through a double barrier resonant tunneling diode whose
transmission peak is narrow and matched to the required injection energy. This 
approach results in a nearly monochromatic injection even at elevated 
temperatures. Accordingly, a high injection efficiency can be maintained 
even for $T$ $>$ 0 K.

This paper is organized as follows. In Section 2, we  calculate the spin-dependent 
contact conductance between a non-ideal metallic ferromagnetic contact and a 
quasi one-dimensional semiconducting wire formed using a combination of mesa etching and 
electrostatic confinement as shown in Fig. 1. This problem is of relevance 
considering the many experimental efforts to realize Spin Field Effect Transistors
 \cite{datta}.

\section{Spin dependent interface conductance}

The structure that we consider is shown in Fig. 1. The ferromagnetic contact that injects electrons into the quantum wire
(which we call the 
source contact)  is quasi one-dimensional, but
the extracting  contact (which we call the drain contact) 
is  a two-dimensional electron gas to which the quantum wire opens up on the right.  
The questions we address are: in the linear mode of operation (i.e., for a small
bias between the source and drain contact) how is the contact resistance of the 
Fe/Sm interface affected by the presence of the Rashba interaction due to the heterostructure
interface formed between materials I (narrow bandgap) and II (wide bandgap) 
in the semiconducting channel? How is that conductance affected by the presence
of a magnetic field applied in the direction of current flow?  The magnetic 
field can be either an external magnetic field or the  stray magnetic field that exists in the vicinity of the Fe/Sm
interface because of the magnetized source contact. In this paper, we consider 
an external magnetic field which has uniform strength along the length of the wire.

In ref. \cite{cahay1}, we  found that the energy dispersion 
relationships (E-$k_x$) for the lowest energy bands in the semiconducting channel
have the general shape shown in Figure 2(a). As was shown in \cite{cahay1}, close
to the bottom of the lowest subband, the E-$k_x$ relationship has a camel-back shape.
Here we will show that this is strictly valid for a magnetic field 
below some critical value $B_c$. Above that threshold, the E-$k_x$ relationships
are as shown in Fig. 2(b) where the camel-back feature of the bottom subband disappears.

If the potential applied to the gate in Fig. 1 is changed, the potential
step between the bottom of the conduction band in the ferromagnetic layer and the 
bottom of the conduction band in the semiconductor quantum wire far from the interface 
($\Delta E_c$) will change.  
As a result, the Fermi level  can be swept from 
below the energy level $E_1$ (bottom of the lower subband) to above 
$E_3$ (bottom of the upper subband).  

Depending on whether the magnetic field is below or above the threshold value $B_c$, the following
two possibilities will occur. Referring to Fig. 2(a) ( B $<$ $B_c$), if we start
with a value of $\Delta E_c$ such that $E_F$ is above $E_3$, there will be initially two 
propagating modes in the semiconductor channel. 
When $E_F$ is in the range $[E_2, E_3 ]$, the 
upper band will become an evanescent channel with a wavenumber which is equal to zero 
when $E_F$ is exactly equal to $E_3$ or $E_2 $, whereas the lower subband is still conducting.
In the energy range $[ E_1, E_2 ]$, the two channels are conducting again and there
should be a rise in conductance until $E_F$ reaches $E_1 $. At that point,
 wavevectors corresponding to the Fermi energy in both subbands will be real, equal and {\it finite}. With a slight
increase in $\Delta E_c$, the Fermi level will fall below $E_1$ and both subbands will
be evanescent. We therefore expect a sharp drop in the conductance as the Fermi level
falls below $E_1$ and the semiconducting channel is completely pinched-off. 

For the case where the magnetic field is above the critical value $B_c$, 
the interface conductance should show a kink as the Fermi level is swept from
above to below the threshold energy $E_3$. Since the upper mode
stays evanescent after $E_F$ drops from $E_3$ to $E_1$ and the lower mode
has a propagating wavevector which gradually shrinks to zero as $E_F$ approaches
$E_1$, the conductance of the interface should smoothly approach zero as the channel
is being pinched-off, contrary to the previous case.

The above discussion pertains to temperature $T$ = 0 K. 
At elevated temperatures, the effect of thermal averaging will smooth out 
any of the abrupt  features in the conductance versus gate voltage characteristics
discussed above. 
At finite temperature and for low value of the bias across the interface 
(linear mode of operation), the conductance of the interface due to the majority 
spin is given by
\begin{equation}
G_{\uparrow} = \frac{e^2}{4 h k_B T} \int_{0}^{\infty} T_{\uparrow}  
sech^2 ( \frac{ E- E_F}{2 k_B T} ) dE,
\end{equation}
where $  T_{\uparrow} $ is the total transmission coefficient of the
majority spin incident from the contact and E is the energy of the electron
incident from the ferromagnetic contact. 
A similar expression is used to calculate the conductance of the minority spin band 
by replacing $ T_{\uparrow} $ by $ T_{\downarrow} $. 
The total conductance of the interface $G_{tot}$ is then equal to 
$ G_{\uparrow} +  G_{\downarrow} $ since the two spin states are orthogonal in the contacts.

In this paper, we investigate the influence of an axial magnetic field
and finite temperature on the spin injection efficiency ($\eta$)
of the Fe/Sm quantum wire interface which is defined as follows
\begin{equation}
\eta = \frac{ I_{\uparrow} -  I_{\downarrow} }{ I_{\uparrow} +  I_{\downarrow} }
 = \frac{ G_{\uparrow} -  G_{\downarrow} }{ G_{\uparrow} +  G_{\downarrow} }.
\end{equation}
where $I_{\uparrow}$ and $I_{\downarrow}$ are the spin polarized currents.

\subsection{Dispersion relations}

The  single 
particle effective-mass Hamiltonian describing the quantum wire subjected to 
an axial magnetic field and Rashba spin orbit interaction is \cite{cahay2}
\begin{equation}
H = {{1}\over{2 m^*}} \left ( {\vec p} + e{\vec A} \right )^2 
+ V_1(y) + V_2(z) 
+  ({g^*}/2) \mu_B {\vec B} \cdot {\vec \sigma} 
+  \frac{{\alpha}_R}{ \hbar} \hat{y}\cdot \left [ {\vec \sigma} \times ( {\vec 
p} + e {\vec A} ) 
\right ]
\end{equation}
where $\hat{y}$ is the unit vector normal to the heterostructure interface in 
Fig. 1, ${\vec A}$ is the vector potential due to the axial magnetic field ${\vec B}$ 
along the channel, 
and  $g^*$ the Land\'{e} factor in the channel \cite{gstar}.
The quantity ${\alpha}_R$ is the Rashba spin-orbit coupling constant 
which varies with the applied potential on the gate \cite{nitta}.  
We will assume that the confining potentials along the y- and z-directions 
are $V_1(y)$ and $V_2(z)$ with the latter being parabolic.

The choice of the Landau gauge ${\vec A}$ = (0, -Bz, 0)  allows us to decouple 
the y-component of the Hamiltonian in (3) from the x-z component.
Accordingly, the two-dimensional Hamiltonian in the plane of 
the channel (x-z plane) is 
\begin{equation}
H_{xz} = {{p_z^2}\over{2 m^*}} + \Delta E_c + {{1}\over{2}}m^* \left ( 
\omega_0^2 + 
\omega_c^2 \right ) z^2 + {{\hbar^2 k_x^2}\over{2 m^*}} +{{\hbar^2 k_R 
k_x}\over{m^*}} \sigma_z + ( g^* /2) \mu_B B {\sigma}_x  - {{\hbar k_R 
p_z}\over{m^*}} \sigma_x
\label{Hamiltonian}
\end{equation}
where $\omega_0$ is the curvature of the confining potential in the z-direction,
$\omega_c$ = $eB/m^*$, ${\mu}_B$ is the Bohr magneton, $k_R = m^* 
\alpha_R/\hbar^2$ is the Rashba wavevector, 
and $\Delta E_c$ is the  potential barrier between the ferromagnet and 
semiconductor. We assume that $\Delta E_c$ includes the effects of the quantum 
confinement in the y-direction. 

For a single moded structure, the energy dispersion relations in the channel can be derived
from Equation (\ref{Hamiltonian}).  
The first five terms of the Hamiltonian in Eq.(\ref{Hamiltonian}) yield shifted parabolic subbands 
with dispersion relations:
\begin{equation}
E_{n, \uparrow} = ( n + 1/2) \hbar \omega + \Delta E_c
+ {{\hbar^2 k_x^2}\over{2 m^*}} + {{\hbar^2 k_R k_x}\over{m^*}}, ~~~
E_{n, \downarrow} = ( n + 1/2) \hbar \omega + \Delta E_c
+ {{\hbar^2 k_x^2}\over{2 m^*}} - {{\hbar^2 k_R k_x}\over{m^*}},
\end{equation}
where $\omega = 
\sqrt{ \omega_0^2 + \omega_c^2 }$. In Eq.(5), the $\uparrow$ and $\downarrow$ 
arrows indicate +z and -z 
polarized spins (eigenstates of the $\sigma_z$ operator) which are split by the 
Rashba effect (fifth term in Equation (4)). These 
subbands have definite spin quantizations axes along +z and -z directions. 

The sixth and seventh terms in Eq.(4) induce a mixing between 
the +z- and -z-polarized spins. The sixth term originates from the magnetic field due to the ferromagnetic 
contacts and the seventh originates from the Rashba effect itself.   The ratio of the sixth and seventh term can be shown to be of the order of 
10$^4$ - 10$^6$ for typical values of the relevant parameters. Therefore, we can neglect the seventh term in comparison 
with the sixth term.

To obtain an analytical expression for the dispersion relation 
corresponding to the first six terms in the Hamiltonian in Eq.(4),  we 
derive a two-band dispersion relation in a truncated Hilbert space considering 
mixing between the two lowest unperturbed subband states (namely the +z and -z spin 
states).  Straightforward diagonalization of the Hamiltonian in Eq.(4) (minus the 
seventh term) in the basis of these two unperturbed states gives the following 
dispersion relations for the two subbands:
\begin{equation}
E_1 (k_x) = {{1}\over{2}} \hbar \omega + \Delta E_c + {{\hbar^2 k_x^2}\over{2 
m^*}}
- \sqrt{ \left ({{\hbar^2 k_R k_x}\over{m^*}} \right )^2 + \left (
\frac{ g^* \mu_B B}{2} \right )^2 },
\label{dispersion1}
\end{equation}
\begin{equation}
E_2 (k_x) = {{1}\over{2}} \hbar \omega + \Delta E_c + {{\hbar^2 k_x^2}\over{2 
m^*}}
+ \sqrt{ \left ( {{\hbar^2 k_R k_x}\over{m^*}} \right )^2 + \left (
\frac{ g^* \mu_B B}{2} \right )^2 }.
\label{dispersion2}
\end{equation}
These dispersion relations are plotted schematically as solid lines in Fig. 2.
$E_1 (k_x) $ refers to the lower subband whereas $E_2 (k_x) $ refers to the upper subband.

The upper subband  $ E_2 (k_x) $ has a global 
minimum at $k_x$ = 0 for all values of B-field and $k_R$.
The location of the minimum is at an energy
\begin{equation}
E_3 = \frac{ {\hbar} \omega }{2} + \Delta E_{c} +  \frac{ | g^{*} |  \mu_{B} B}{2}
\end{equation}
indicated on Fig.2. This energy corresponds to the upper Zeeman energy level in the
semiconductor channel.

The lower subband  $ E_1 (k_x) $ has
a camel-back shape as long as the external magnetic field $B$ is
less than some critical value $B_c$. The latter can be obtained by setting the derivative of the $E-k_x$ 
relationship equal to zero. The local maximum is located at $k_x = 0$ whereas
two local minima are located at 
\begin{equation}
k_{min} = \pm k_R \sqrt{ 1 - \frac{1}{16} ( \frac{ g^{*} \mu_{B} B}{ {\delta}_R })^2 }.
\end{equation}
The quantity $k_{min}$ is real
only if the external magnetic field $B$ is less than the critical value
$B_{c} $ given by
\begin{equation}
| g^{*} | {\mu}_B B_{c} =  4  {\delta}_R ,
\end{equation}
i.e., when the Zeeman splitting is four times the Rashba spin splitting energy
$ {\delta}_R $:
\begin{equation}
{\delta}_R =  \frac{  {\hbar}^2 {k_R}^2 }{ 2 m^*}.
\end{equation}
The value of $ E_1 (k_x) $ at $k_x = 0 $ corresponds to the lower Zeeman
energy level in the semiconductor channel and is given by
\begin{equation}
E_2 = \frac{ {\hbar} \omega }{2} + \Delta E_{c} - \frac{| g^{*} | \mu_{B} B}{2},
\end{equation}
whereas the value of the energy at the minima $k_{min}$ is given by
\begin{equation}
E_1 = \frac{ {\hbar} \omega }{2} + \Delta E_{c} 
- [ {\delta}_R + \frac{ ({g^{*} \mu_{B} B})^2 }{ 16 {\delta}_R }  ].
\end{equation}

From Equations (\ref{dispersion1} - \ref{dispersion2}), we find that an 
electron with  energy $E$ has wavevectors in the two 
bands  given by
\begin{equation}
k_{x,1} = \frac{1}{\hbar} \sqrt{ 2m^*  ( \frac{ A + \sqrt{A^2 - 4C}  }{2}) }, ~~~ 
k_{x,2} = \frac{1}{\hbar} \sqrt{ 2m^*  ( \frac{ A - \sqrt{A^2 - 4C}  }{2}) }
\end{equation}
where
\begin{equation}
A = 2 (E - \frac{ \hbar \omega}{2} - \Delta E_c ) + 4 {\delta}_{R}, ~~~
C = ( E - \frac{ \hbar \omega}{2} - \Delta E_c )^2 - {\beta}^2, 
~~~ \beta =  | g^* | \mu_B B/2.
\end{equation}

It is obvious that when the external magnetic field $B$ $< B_c$, in the energy range between $E_2$ and $E_3$, C is negative and $k_{x,1}$ is real
while  $k_{x,2}$ is purely imaginary. This means that there is only one propagating channel 
in this energy range (the other channel is evanescent). On the other hand, if $E$ is between $E_1$ and $E_2$, or if $E$ is larger than $E_3$,
then both $k_{x,1}$ and $k_{x,2}$ are real, so that there are two propagating channels.

If B $>$ $B_c$, $E_1$ is equal to $E_2 $ since the bottom subband looses it camel-back shape
and has only a minimum at $k_x = 0$.  In that case, if $E$ is between $E_1$ and $E_3$, then
A is negative and  $k_{x,2}$ is purely imaginary while  $k_{x,1}$ is real. Thus, there is only one propagating 
channel. If $E$ $> E_3$,
then both A and C are positive and $k_{x,2}$, $k_{x,1}$ are both real, so that there 
are two propagating channels.

If $E < E_1$, then there is no propagating channel and the conductance drops 
sharply towards zero.

The eigenspinors corresponding to the eigenvalues given in Eqns.(6-7)
have the explicit forms \cite{cahay2}
\begin{eqnarray}
\left [ \begin{array}{c}
             C_1 (k_{x,1})\\
             {C^{'}}_1 (k_{x,1})\\
             \end{array}   \right ]
& = &
 \left [ \begin{array}{c}
\alpha (k_{x,1})/\gamma (k_{x,1})\\
\beta/\gamma (k_{x,1}) \\
\end{array} \right ] \nonumber \\
\left [ \begin{array}{c}
             C_2 (k_{x,2})\\
             {C^{'}}_2 (k_{x,2})\\
             \end{array}   \right ]
& = &
 \left [ \begin{array}{c}
- \beta/\gamma (k_{x,2}) \\
\alpha (k_{x,2}) /\gamma (k_{x,2})\\
\end{array} \right ]
\end{eqnarray}
where the quantities $\alpha$ and $\gamma$ are  given 
by
\begin{equation}
\alpha (k_{x})  =  {{\hbar^2 k_R k_x}\over{m^*}} + \sqrt{ \left ( {{\hbar^2 
k_R 
k_x}\over{m^*}} \right )^2 + {\beta}^2 }, ~~~
\gamma ( k_x ) =  \sqrt{ \alpha^2 + \beta^2}.
\end{equation}

As pointed out in ref. \cite{cahay1}, these eigenspinors are not 
+z-polarized state $\left 
[ \begin{array}{cc}
1 & 0 
\end{array} \right ]^{\dagger}
$,  or -z-polarized state $\left [ \begin{array}{cc}
0 & 1 
\end{array} \right ]^{\dagger} 
$ if the magnetic field $B \neq 0$. Thus, the magnetic field mixes spins
and the +z or -z polarized states are no longer eigenstates in the semiconducting channel.
Equations (16) also show that the spin quantization axis (eigenspinor) in any subband 
is not fixed and strongly depends on the wavevector $k_x$. Thus, an electron 
entering the semiconductor channel from the left ferromagnetic contact with +x-polarized 
spin, will not couple {\it equally} to +z and -z states. The relative coupling will depend on the 
electron's energy. 

\subsection{The transport problem across the interface:} We model the ferromagnetic contact by 
the Stoner-Wohlfarth model. The magnetization of the contact is assumed to be 
along the x-direction so that the 
majority carriers are +x-polarized electrons (as in ref. \cite{datta}) and 
minority carriers are -x-polarized. Their bands are offset by an 
exchange splitting energy $\Delta$ (Fig. 2).  In the expression of the Hamiltonian 
in Eq.(4) we add an extra potential energy term
$V_I(x)$ to represent an interfacial potential
barrier between the metallic ferromagnetic contact and the semiconducting channel.
Following Sch\"apers et. al. \cite{schapers2}, we model the interface barrier as a delta-scatterer:
\begin{equation}
V_I(x) = V_0 \delta (x),
\end{equation}
where $ V_0 $ is the strength of the scattering potential.

In the semiconducting quasi one-dimensional channel( $ x > 0$), the
x-component of the wavefunction at a position $x$ along  
the channel is given by
\begin{eqnarray}
{\psi}_{II} (x) & = &
T_1
 \left [ \begin{array}{c}
             C_1(k_{x,1})\\
             C'_1 (k_{x,1})\\
             \end{array}   \right ]
             e^{i k_{x,1} x} 
             +           
T_2
 \left [ \begin{array}{c}
             C_2(k_{x,2})\\
             C'_2 (k_{x,2})\\
             \end{array}   \right ]
             e^{i k_{x,2} x} ,
\label{wavefunction}
\end{eqnarray}
where $T_1$ and $T_2$ are the transmission amplitudes into the two subbands given by
Equations (\ref{dispersion1}) and (\ref{dispersion2}).

For an electron in the {\it majority spin band} in the left ferromagnetic contact (region I; $x < 
0$), 
the electron is spin polarized in the
$\left [ \begin{array}{cc}
             1, 1
             \end{array}
             \right ]^{\dagger} $ 
subband and the x-component of the wavefunction 
is given by
\begin{eqnarray}
{\psi}^{majority}_I (x) & = &
 \frac{1}{\sqrt{2}} \left [ \begin{array}{c}
             1\\
             1\\
             \end{array}   \right ]
             e^{i {k_{x}}^u x}
             +
  \frac{ R_1}{\sqrt{2}} \left [ \begin{array}{c}
             1\\
             1\\
             \end{array}   \right ]
             e^{-i {k_{x}}^u x} 
             +
 \frac{R_2}{\sqrt{2}} \left [ \begin{array}{c}
              1\\
             -1\\
             \end{array}   \right ]
             e^{-i {k_{x}}^d  x},
\end{eqnarray}
where $R_1$ is the reflection amplitude into the +x-polarized band and $R_2$ is 
the 
reflection amplitude in the -x-polarized band.

The wavevectors
\begin{equation}
k_x^u  = \frac{1}{ \hbar} \sqrt{2 m_0 E_F }, ~~~ k_x^d = \frac{1}{ \hbar} \sqrt{2 
m_0 (E_F - \Delta)},
\end{equation}
are the x components of the wavevectors in the +x (majority spin) 
and -x-polarized (minority spin) energy bands in the ferromagnet, respectively.

For an electron in the {\it minority spin band} in the left 
ferromagnetic contact (region I; $x < 0$), 
the electron is spin polarized in the
$\left [ \begin{array}{cc}
             1, -1
             \end{array}
             \right ]^{\dagger} $ 
subband and the x-component of the wavefunction 
is given by
\begin{eqnarray}
{\psi}^{minority}_I (x) & = &
 \frac{1}{\sqrt{2}} \left [ \begin{array}{c}
             1\\
             -1\\
             \end{array}   \right ]
             e^{i {k_{x}}^u x}
             +
  \frac{ R'_1}{\sqrt{2}} \left [ \begin{array}{c}
             1\\
             -1\\
             \end{array}   \right ]
             e^{-i {k_{x}}^u x} 
             +
 \frac{R'_2}{\sqrt{2}} \left [ \begin{array}{c}
              1\\
             1\\
             \end{array}   \right ]
             e^{-i {k_{x}}^d  x},
\end{eqnarray}
where $R'_1$ is the reflection amplitude into the -x-polarized band and $R'_2$ is 
the 
reflection amplitude in the +x-polarized band.

The four unknowns ($R_1$,$R_2$,$T_1$,$T_2$) or ($R'_1$,$R'_2$,$T_1$,$T_2$) are 
 solutions of a 4x4 system of equations obtained by 
enforcing continuity of the wavefunction and the quantity
\begin{equation}
\mu \frac{d \psi }{dx} (-\epsilon)  + \frac{2 {m_s}^* V_0}{ {\hbar}^2 } \psi (0) =
\frac{d \psi }{dx} (+\epsilon)  +  i k_R (+\epsilon) {\sigma}_z \psi
(+\epsilon),
\label{bc1}
\end{equation}
where $\mu = \frac{ {m_s}^* }{
{m_f}^* } $ and $ {m_s}^* $ (${m_f}^*$)
is the effective mass in the semiconductor (ferromagnet). 
Here, we have made use of the fact that $\alpha_R$ (and therefore
$k_R$) is zero in the ferromagnetic contacts so that a term containing $k_R
(-\epsilon)$ does not appear in Eq.(23). 
This last equation ensures continuity of the current density at the
ferromagnetic contact/semiconductor interface \cite{cahay2}.

\subsection{The interface conductance}
 
At $T$ = 0 K, the conductance of the interface is calculated using the Landauer formula, i.e.,
the sum over the majority and minority spin subbands
of $e^2 /h $ times the transmission coefficient across the interface for each subband.
For each subband, the transmission coefficient is the ratio of the current
density in the semiconducting channel divided by the current density of the incident beam in the ferromagnetic contact.

The transmitted current density in the semiconductor channel is found by plugging Eq.(19)
into the current density expression
\begin{equation}
J_{x} = \frac{\hbar}{2m^{\ast}i}(\Psi^{\dagger}\frac{d\Psi}{dx}
-\frac{d\Psi}{dx}^{\dagger}\Psi)
+\frac{\alpha_{R}}{\hbar}(\Psi^{\dagger}\sigma_{z}\Psi)
\end{equation} 

For electron with incident energy within the energy range [$E_2, E_3 $] where one of
the mode is evanescent in the channel, this leads to 
\begin{eqnarray}
J_{x}&=&\frac{\hbar}{m^{\ast}}\left[|T_1|^2 ( k_{x,1}+k_R(|C_1|^2-|C_1'|^2) )
+|T_2|^2k_R(|C_2|^2 -|C_2'|^2)\right]\nonumber\\
&+&\frac{2 \hbar k_R}{m^{\ast}} Re\left[T_2 T_1^{\star} (C_2 C_1^{\star} 
- C_2' C_1'^{\star})\right]\nonumber\\
&+&\frac{\hbar}{m^{\ast}} 
Re\left[T_2T_1^{\star} ( C_2 C_1^{\star} +  C_2'C_1'^{\star} ) (i\kappa+k_{x,1})\right],
\end{eqnarray}
where the $\star$ stands for complex conjugate.

At $T$ = 0 K, the conductance of the interface due to the majority spin band is then given 
by
\begin{equation}
G_{\uparrow} = \frac{e^2}{h} T_{\uparrow}
= \frac{e^2}{h}( {J_x} / {J_x}^{inc} )
\end{equation}
where ${J_x}^{inc} = \hbar {k_x}^u /m_o $. A similar calculation yields
$G_{\downarrow}$. As a check of our numerical simulations, we also compute
the total reflection coefficient for electrons incident in the majority spin band
\begin{equation}
R_{\uparrow}  = |R_1 |^2  + \frac{k_x^d}{k_x^u}|R_2 |^2 ,
\end{equation}
and electrons incident in the minority spin band
\begin{equation}
R_{\downarrow} = \frac{k_x^u}{k_x^d} |R_1 |^2  + |R_2 |^2 ,
\end{equation}
and check numerically that $R_{\uparrow} + R_{\downarrow} + T_{\uparrow} + T_{\downarrow} = 1$.

The last term in Eq.(25) is proportional to the overlap factor
$( C_2 C_1^{\star} +  C_2'C_1'^{\star} )$ between the two eigenspinors in the channel 
and is non-zero in the presence of an axial magnetic field which makes the two spinors
(at a given energy) non-orthogonal. 
This is an important feature which leads to the presence of antiresonances in 
the transmission coefficient of electrons incident in the minority spin band, 
as discussed in the next section. Since the two spinors are non-orthogonal only
if  there is (i) an axial magnetic field, and (ii)  a spin orbit interaction
in the semiconductor, it is obvious these two conditions are necessary for the 
anti-resonances to occur. Accordingly, the effect reported in this paper, namely 
100\% spin injection efficiency caused by anti-resonances, requires a magnetic 
field and spin-orbit interaction.

\section{Results and discussions}

We consider a Fe/Sm quantum wire interface with the geometry shown in Fig. 1
and with the parameters listed in Table I.  
The electrostatic confinement along the z-direction is 
assumed to be parabolic with $  \hbar \omega$ = 10 meV.

For simplicity, we assume that the Rashba spin-orbit coupling constant
${\alpha}_R $ and the strength $V_0$ of the Schottky barrier height at the
Fe/Sm quantum wire interface are independent of the gate potential.
For an interface with the parameters listed in Table I, the critical magnetic field $B_c$ below which 
the E-$k_x$ relationship for the lowest propagating mode in the semiconducting channel 
has a camelback shape is equal to 0.07 Tesla. Figure 3 is a plot of the magnetic 
field dependence of the threshold energies $(E_1 , E_2, E_3) $.

Figure 4 is a plot of the wavevectors in the two subbands corresponding to the Fermi
energy as a function of the electron incident energy
for an external magnetic field of 0.02 T (which is below)
and 0.2 T (which is above) the critical value $B_c$ (0.07 T).  
Figure 4(a) shows that the top subband
becomes evanescent when the Fermi level in the contact falls within the energy
range [$E_2, E_3 ]$. In this energy range, $ k_{x,2} $ is purely
imaginary and we have plotted in Fig.4 the quantity $\kappa$ such that $k_{x,2}$ = i $\kappa$.
Note that $\kappa$ is exactly zero at $E_2$ and $E_3 $ as can easily be shown using the
 equations derived in Section 1.1. Furthermore, $\kappa$ reaches
a maximum in the energy range [$E_2, E_3 ]$ which can be shown to be located
at an energy $E^*$ given by
\begin{equation}
E^* = \frac{ {\hbar} \omega }{2} + \Delta E_{c} - \frac{ {\beta}^2 }{ 4 {\delta}_R }.
\end{equation}
When $B > B_c$ as in Fig.4(b),  $ k_{x,2} $ is purely
imaginary in the energy range [$E_1, E_3 ]$ 
and we also plot the value of $\kappa$ such that $k_{x,2}$ = i $\kappa$
in that interval.

Figures 5 and 6 show plots of the conductance (in units of $e^2$/h) variation 
with ${\Delta} E_c$ for the case where the longitudinal magnetic
field is equal to 0.02 T ($< B_c$) and 0.2 T ($> B_c$), respectively.
Figure 5 shows an abrupt drop in the conductance from a finite value down to zero as the Fermi level
in the contact is brought below the threshold energy $E_1$ shown in Fig.2(a). As discussed
in Section 1, the sudden drop is related to the fact that, right at the threshold energy, 
the wavevector of the two free propagating modes in the semiconductor channel are equal and non-zero.
On the other hand, the conductance goes down {\it smoothly} to zero
when $B > B_c $ (Fig. 6) since the wavevector of the only propagating mode vanishes when $E_F$ crosses
the threshold energy $E_1$ in Fig. 2(b).

Figures 5 and 6 show that the conductance of the interface depends strongly on the strength
of the Schottky barrier height $V_0$, with larger values
leading to a smaller conductance.
When the magnetic field is above $B_c$ as in Fig.6, there is a kink in the conductance
appearing at the energy where the upper subband becomes evanescent (Zeeman energy $E_3$ in Fig.2(b)).
When $B < B_c$, the upper subband is only evanescent when the Fermi energy is
within the energy range [$E_2, E_3$] (See Fig.4(a)). The conductance reaches
a minimum at the edges of this energy range and shows another kink within the interval
at the energy $E^*$ given by Eq.(29). 
Also shown in Figures 5 and 6 is $ G_{tot}$ versus $ \Delta E_c $
when there is no axial magnetic field in the channel (for the case $V_0$ = 0 eV $\AA$).
In this case, $ G_{tot}$ is a smooth function of $ \Delta E_c $ and goes down
to zero smoothly at threshold. This feature is found to be independent of the
Schottky barrier strength at the interface. In the absence of an externally applied magnetic
field, the presence of kinks in the interface conductance at low temperature highlights
the non-trivial effect that a stray magnetic field in the channel has.

The kink in Fig.5 and the local minima in $G_{tot}$ in Fig. 6 are due to antiresonances
in the transmission coefficient of the minority spin band when the Fermi level in the ferromagnetic
contact is lined up with the Zeeman energy levels in the semiconductor channel, as illustrated
in Figures 7 and 8, for B $< B_c$ and B $ > B_c $, respectively.
At zero temperature, these antiresonances lead to 
a {\it perfect}  ($\eta$ = 100 $\%$) spin injection efficiency, as illustrated in Figs. 10 and 11
for the case of B $< B_c$ and B $ > B_c $, respectively. One important feature is that
this perfect spin injection efficiency is {\it independent}
of the value of the Schottky barrier height. This means that no Schottky barrier 
or tunnel barrier is necessary to implement the 100\% spin injection efficiency.
Perfect spin filtering (or 100\% spin injection efficiency, i.e. $\eta$ = 1)  
occurs when the Fermi level $E_F $ in the ferromagnetic contact coincides with $E_2$ or $E_3$  
levels in the channel when B $< B_c$ and when $E_F$ coincides with $E_3$
when B $> B_c$.

\subsection{Importance of the magnetic field}

For both $B < B_c $ and $B > B_c $,
$[  C_2(k_{x,2}), C'_2 (k_{x,2} ]^{T}$ (where T stands for transpose)
reduces to $ \frac{1}{\sqrt{2}} [1,-1]$ when the Fermi level in the contact
coincides with $E_2$ or $E_3$.
On the other hand, $[  C_1(k_{x,1}), C'_1 (k_{x,1} ]^{T}$ is not equal to
$ \frac{1}{\sqrt{2}} [1,1]$ at either $E_2 $ or $E_3$, when $B < B_c$, but it is equal
to $ \frac{1}{\sqrt{2}} [1,1]$ at $E_2 $ when $ B > B_c$.

When $B < B_c $, the overlap factor in Eq.(24) is  non-zero when $E_F = E_2$ or $E_3$.
In this case, it can be shown that both $T_1$ and $T_2$ in Eq.(25) are non-zero
at these incident energies for electrons incident in both majority and minority spin bands.
Furthermore, in  the case of minority spin band, the contribution from the overlap
term in Eq.(25) exactly cancels the other contributions leading to zero transmitted current
in the channel \cite{unpublished}. 

When $B > B_c $, $T_1 $ = $T_2 $ = 0 for both minority and majority spin bands incident 
from the contact when $E_F = E_2 $. In that case the 4x4 system of equations
reduces to two 2x2 systems of equations with no mixing between the two subbands in the channel.

%

The maximum spin injection efficiency is much less than 100\% and virtually independent 
of $ \Delta E_{c} $ when the axial magnetic is assumed to be {\it zero}, as illustrated by the
dashed line in Figures 9 and 10. In that case, the spin subbands are orthogonal
in the channel and there is no antiresonance in the transmission coefficient
of the minority spins. Thus, the axial magnetic field is critical to ensure perfect 
spin filtering and 100 $\%$ spin injection efficiency!

\subsection{Role of tunnel barrier}

Another important feature in Figures 10 and 11 is that over a given range of $ \Delta E_{c} $,
(or, equivalently, over a range of injection energies), the spin injection efficiency is larger  if
the Schottky barrier height at the interface is larger. This means that although no barrier is required 
for perfect spin filtering  at $T$ = 0 K, a barrier is beneficial 
at $T$ $>$ 0K when thermal smearing causes a spread in the injection energy. 
The presence of a taller barrier leads
 to improved spin filtering over a {\it range of energies}. 
This is consistent with the result of ref. \cite{rashba}. The difference with ref. \cite{rashba}
is that we consider {\it ballistic} transport whereas ref. \cite{rashba} considered 
{\it diffusive} transport. Thus, we have extended the basic finding of ref. \cite{rashba}
to the regime of ballistic transport.

\subsection{Transconductance}

Since the quantum wire will be typically formed using a split-gate, a more negative
voltage on the gate will lead to an increase of $ \Delta E_{c} $. Fig.5
shows that, at very low temperature when B $< B_c$, there are regimes of operation
where the transconductance, $ d G_{tot}/ d V_{G} $, can be either positive or negative.
In contrast, it is always positive when B $ > B_c$, as shown in Fig. 6.
Since the regime of negative transconductance occurs over a very narrow range of $ \Delta E_{c} $,
they can only be observed at very low temperature. This is illustrated in Fig.9(a) where
$G_{tot}$ is plotted versus $ \Delta E_{c} $ as a function of temperature when  B $< B_c$.
Figure 9(b) also shows that the kink in $G_{tot}$, appearing  when B $ > B_c$, is rapidly smoothed
out as the temperature increases. Even though not shown here, the curves for
$G_{tot}$ in Figures 9(a) and 9(b) at non-zero temperature were found to smoothly reach
zero at larger values of $\Delta E_c$.

\subsection{Temperature dependence of the spin injection efficiency}

The spin injection efficiency decreases rapidly as the temperature increases, especially 
when the axial magnetic field is below the critical magnetic field value $B_c$.
For devices with GaAs, InAs, or InSb semiconducting channels
operating with a value of the Rashba spin-orbit coupling constant 
${\alpha}_R$ equal to $10^{-11} $ eV cm, $B_c$ is equal to  6.89, 0.07, and 0.012 Tesla,
respectively. Figures 12 and 13 show that
the maximum spin injection efficiency rapidly degrades as the temperature
rises. The degradation is faster when the axial magnetic field is below 
the critical value $B_c$. For instance, at T = 0.1 K, the spin injection efficiency
is still around 82 $\%$ when B $> B_c$ whereas it is down to 12 $\%$ when
B $< B_c $.

This can be understood as follows. In the Landauer-B\"{u}ttiker formalism,
the linear response conductance is given by Eq.(1) and is a convolution
over energy of the transmission coefficient and the bell shape function ($sech^2$)
centered just a few $k_B T$ around the Fermi level.
As illustrated in Figures 9(a) and 9(b), the transmission
coefficient of the minority spin band in the energy range $[ E_1, E_F + 2 k_B T ] $ is smaller
when B $> B_c$ compared to the case when B $< B_c $. Therefore, the contribution to the conductance
due to minority spins will be less when B $> B_c$. This leads to a less rapid decrease
of the spin injection efficiency with temperature when B $> B_c$.

An estimate of the temperature $T^*$ at which the spin injection efficiency
will decrease substantially can be obtained from the equation,
\begin{equation}
4 k_B T^* = | g^* | {\mu}_B B,
\end{equation}
which equates the approximate width of the bell shaped function ($sech^2$) in Eq.(1)
to the splitting between the two Zeeman energy states in the channel. Table II
lists $T^*$ for different semiconductor materials and two different values of the
axial magnetic field.
This table indicates that a larger spin injection efficiency should be sustainable
at higher temperature in devices based on InSb channels. Table II also indicates
that a rapid decrease of the spin injection efficiency is expected at a
temperature of 0.05 K and 0.5 K when B $< B_c $ and B $> B_c$, respectively, 
in agreement with the plots shown in Figures 12 and 13.

\subsection{Structure for maintaining a large spin injection efficiency at elevated 
temperatures}

It is obvious that the degradation of the spin injection efficiency at elevated 
temperatures is caused by the thermal smearing of the injection energy around the
critical value. The remedy
to this problem is to interpose a double barrier resonant tunneling diode
between the ferromagnet and the semiconductor quantum wire as shown in Fig.14.
The barrier heights and widths are chosen to be sufficiently large so that there is a 
narrow transmission peak at the resonant energy. All we have to do is design the 
double barrier structure such that the resonance energy is matched to the critical 
injection energy needed to achieve a 100\% spin injection efficiency. In that case,
a nearly monochromatic beam of electrons is incident on the semiconductor structure
at the critical energy and this will lead to a high spin injection energy, even at 
relatively elevated temperatures.

\section{Conclusions}

We have shown that at $T$ = 0 K, 100\% spin injection efficiency is possible across the interface 
of a non-ideal metallic ferromagnet and a semiconductor quantum wire, without a 
tunnel barrier, provided we inject carriers with certain  energies. Additional  
requirements are that there must be an axial magnetic field and a spin orbit interaction in the 
quantum wire. We have also described some approaches to retaining a high spin injection 
efficiency at elevated temperatures by designing appropriate structures and choosing 
appropriate semiconductor materials.

The work at Virginia was supported by the Air Force Office of Scientific Research under 
grant FA9550-04-1-0261.
\newpage

\vskip .1in
\begin{center}
{\bf Table I: Parameters of Fe/InAs interface}
\end{center}
\begin{center}
\begin{table}[h]
\centering
\begin{tabular}{cc} \hline
Fermi Energy $E_F$ (eV) & 4.2         \\
Exchange splitting energy ${\Delta}$ (eV)  &  3.46       \\
Rashba spin-orbit coupling constant ${\alpha}_R$ ($10^{-11}$ eVcm)  &  1.    \\
Lande Factor $g^*$ \cite{papp}  &  -14.9   \\
Effective mass ${m_f}^* $ in Fe contact ($m_0$)  &   1.     \\
Effective mass ${m_s}^* $ in InAs channel ($m_0$) 
\cite{papp}  &   0.023     \\ \hline\hline
\end{tabular}
\end{table}
\end{center}

\newpage

\vskip .1in
\begin{center}
{\bf Table II: Critical temperature $T^*$ (Kelvin) \\
for different semiconductor channels}
\end{center}
\begin{center}
\begin{table}[h]
\centering
\begin{tabular}{cccc} \hline
Axial magnetic field (Tesla) & GaAs    & InAs    & InSb   \\
0.02                         & 0.0015  & 0.05    & 0.17   \\ 
0.2                          & 0.015   & 0.5     & 1.73   \\ \hline\hline
\end{tabular}
\end{table}
\end{center}

%
%

%

\newpage
\
\parindent 0cm

\begin{center}
{\bf Figure Captions}
\end{center}

\vskip .1in
{\bf Figure 1}: Ferromagnetic/semiconducting quantum wire contact.
(a) Top view of the structure showing the ferromagnetic contact with its
magnetization in the direction of current flow in the quantum wire. An external
magnetic field is applied along the direction of the wire. The conductance
of the wire is measured between the ferromagnetic source and a (non-magnetic) drain contact located
in the two-dimensional electron gas on the right. (b) Cross-sectional view of the
structure showing as a thick dashed line the quasi one-dimensional electron gas in the wire.

\vskip .1in
{\bf Figure 2}: Energy band diagram across the ferromagnetic contact/semiconductor quantum 
wire interface.  
We use a Stoner-Wohlfarth model for the ferromagnetic contact.
$\Delta$ is the exchange splitting energy in the contacts. $\Delta E_c$ is the height of the
potential barrier between the energy band bottoms of the semiconductor and the ferromagnetic contacts.
Also shown are the energy dispersion relationships on both sides
of the interface for the case of an axial magnetic field (along the x-direction of current flow
shown in Fig.1) below (top figure) or above (bottom figure) the critical magnetic field
value discussed in the text.

\vskip .1in
{\bf Figure 3}: Magnetic field dependence of the energies
$E_1$, $E_2$, and $E_3$ (from bottom to top) versus magnetic field for a quantum wire
with the parameters listed in Table I. For $B > B_c $ (0.07 T in our simulations), $E_1 = E_2$.

\vskip .1in
{\bf Figure 4}: Energy dependence of the wavevectors
in the two lower subbands for a magnetic field of B = 0.02 T which is below
the critical value $B_c$ = 0.07 T (top figure) and for B = 0.2 T which is above  $ B_c$ (bottom figure).
The interface parameters are listed in Table I. The threshold
energies $E_1$, $E_2$, and $E_3$ discussed in section II are indicated.
In the energy range where the upper subband becomes evanescent, we have plotted
$\kappa$, where $k_{x,2}$ = i $\kappa$ with $k_{x,2}$ given by Eq.(14).
The wavevectors are plotted in units of the Rashba wavevector $k_R$.

\newpage
{\bf Figure 5}: Zero temperature variation of the conductance (in units of $e^2 /h$)
of a ferromagnetic contact/semiconducting
nanowire as a function of $\Delta E_c$ for a magnetic field B = 0.02 T which is below the critical
value $B_c$ given by Eq.(10).
The different full lines correspond to different values of $V_0$ characterizing
the strength of the Schottky barrier height at the ferromagnet/semiconductor interface.
From top to bottom, $V_0 $ = 0, 1, and 2 eV$\AA$, respectively.
Also shown as a dashed line is $G_{tot}$ versus $\Delta E_c$ neglecting the axial magnetic
field in the channel for $V_0 $ = 0 eV$\AA$.

\vskip .1in
{\bf Figure 6}: Same as Fig.5 for an axial magnetic field of
0.2 Tesla.

\vskip .1in
{\bf Figure 7}: Energy dependence of the total transmission coefficient of minority 
($ T_{\downarrow}$, full lines) and majority ($ T_{\uparrow}$, dashed lines)  
spin bands across
the ferromagnetic contact/semicon-\\
ductor quantum wire interface over the
energy range [ $E_1, E_F + 2 k_B T ]$ for T = 2K. The axial
magnetic field is equal to 0.02 T. 
For each set of curves, the strength of the Schottky barrier height $V_0 $
from top to bottom is equal to 0, 1, and 2 eV$\AA$, respectively.

\vskip .1in
{\bf Figure 8}: Same as Fig.7 for an axial magnetic field of 0.2 T.

\vskip .1in
{\bf Figure 9}: Conductance (in units of $e^2 /h$)
of a ferromagnetic contact/semiconducting quantum wire as a function of $\Delta E_c$
and temperature. The strength $V_0 $ of the Schottky barrier height
at the ferromagnet/semiconductor interface is set equal to 1 eV$\AA$.
The top and bottom figures are for an axial magnetic field in the quantum wire
equal to 0.02 and 0.2 Tesla, respectively.

\vskip .1in
{\bf Figure 10:} Zero temperature variation of the spin injection efficiency
of a ferromagnetic contact/semiconducting
quantum wire interface as a function of $\Delta E_c$ 
for a magnetic field B = 0.02 T.
The different full lines correspond to different values of $V_0$ characterizing the
strength of the Schottky barrier height at the ferromagnet/semiconductor interface.
The dashed lines are the spin injection efficiency calculated with no axial magnetic
field in the channel. For both sets of curves,
$V_0 $ = 0, 1, and 2 eV$\AA$, from bottom to top.

\vskip .1in
{\bf Figure 11}: Same as Fig.10 for an axial magnetic field B = 0.2 T.

\vskip .1in
{\bf Figure 12}: Temperature variation of the spin injection efficiency
of a ferromagnetic contact/semiconducting quantum wire interface
as a function of $\Delta E_c$ for a magnetic field B = 0.2 T.
From top to bottom, the different curves correspond to
a temperature of 0, 0.1, 0.5, 1, and 2 K, respectively.
The strength of the Schottky barrier height $V_0 $ at the ferromagnet/semiconductor
interface is set equal to 1 eV$\AA$.

\vskip .1in
{\bf Figure 13}: Same as Fig.12 for an axial magnetic field of 0.2 T.

\vskip .1in
{\bf Figure 14}: Proposed double barrier resonant tunneling heterostructure to maintain efficient
spin injection at elevated temperatures.

\newpage
\
\vskip 1.2in
\begin{figure}[h]
\centerline{\psfig{figure=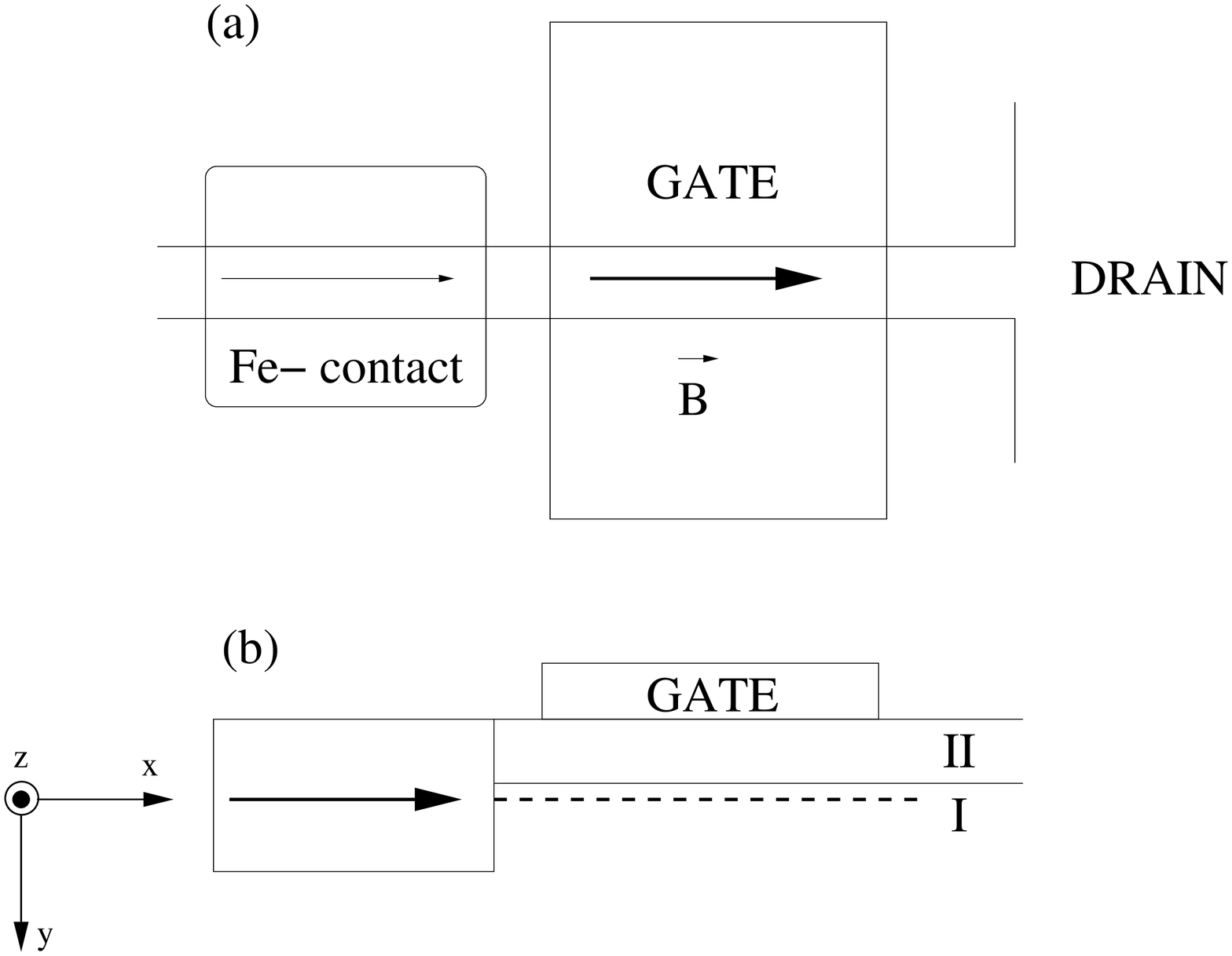,height=4.5in,width=6in}}
\vskip .5in
\begin{center}
Figure 1
\end{center}
\end{figure}

\newpage
\
\vskip .1in
\begin{figure}[h]
\centerline{\psfig{figure=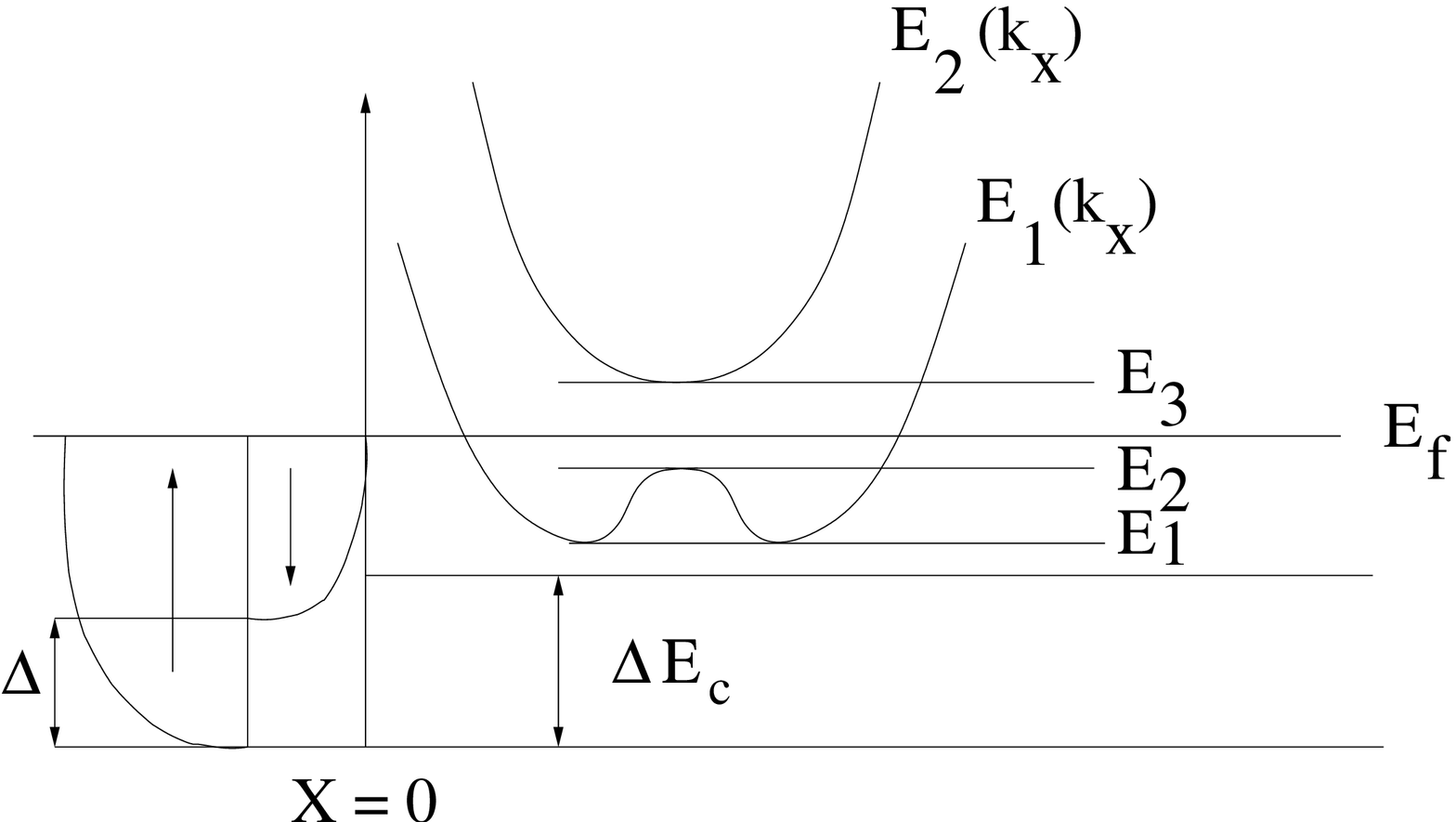,height=2.3in,width=3.5in}}
\end{figure}
\
\vskip .1in
\begin{figure}[h]
\centerline{\psfig{figure=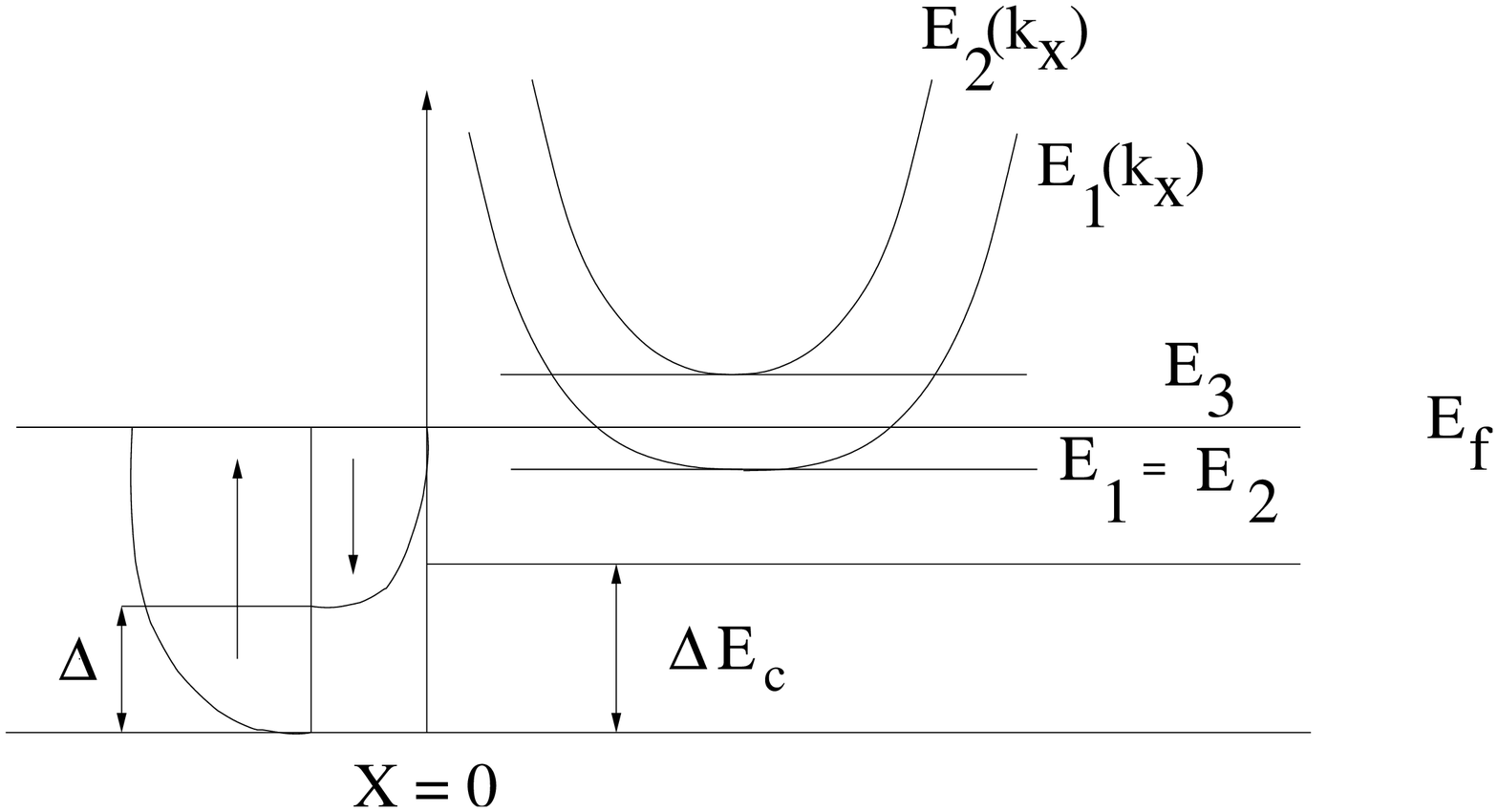,height=2.3in,width=3.5in}}
\vskip .5in
\begin{center}
Figure 2
\end{center}
\end{figure}

\newpage
\
\vskip .5in
\begin{figure}[h]
\centerline{\psfig{figure=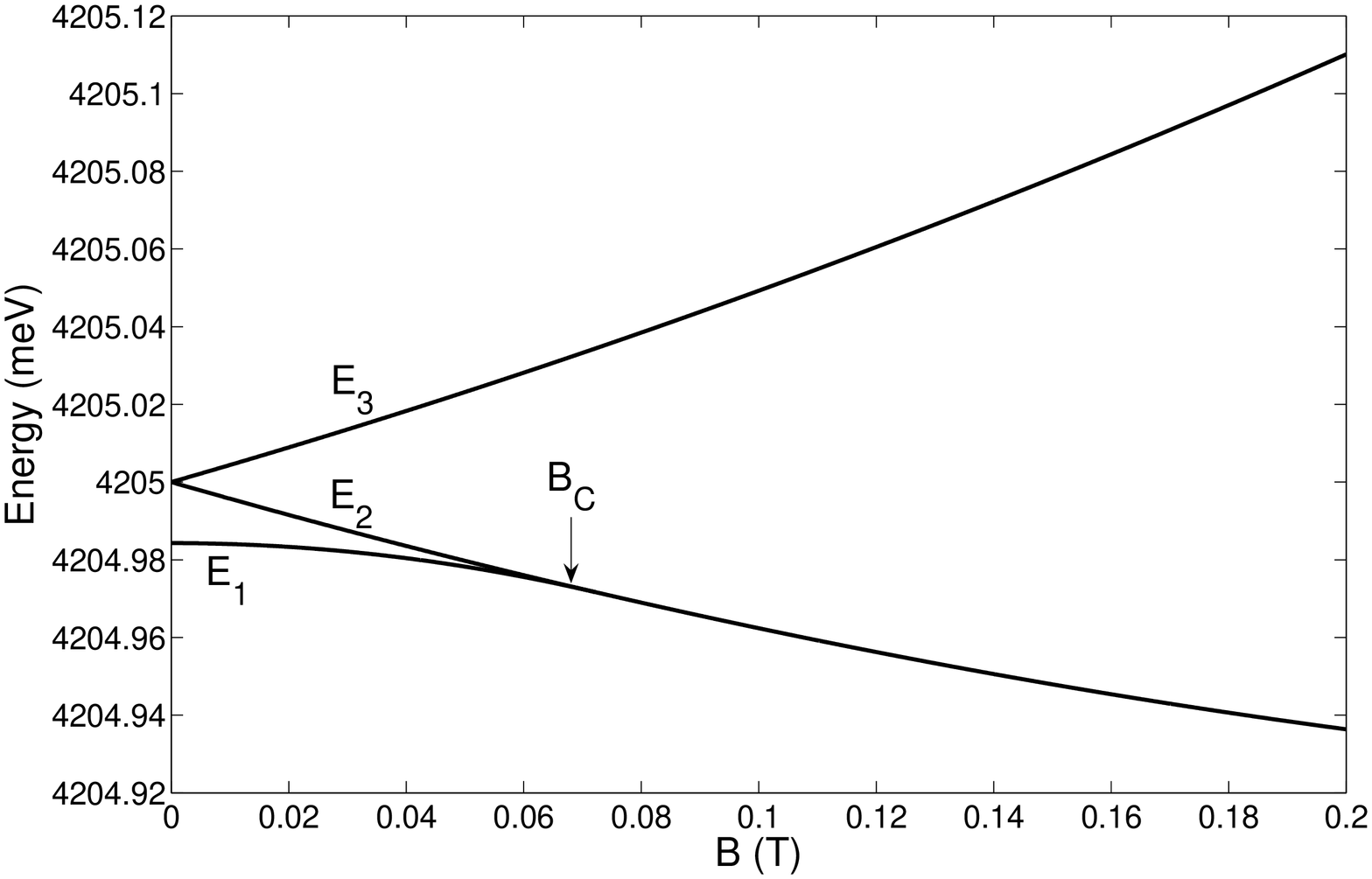,height=5.5in,width=6in,angle=-90}}
\vskip .5in
\begin{center}
Figure 3
\end{center}
\end{figure}

\newpage
\
\vskip 0.5in
\begin{figure}[h]
\centerline{\psfig{figure=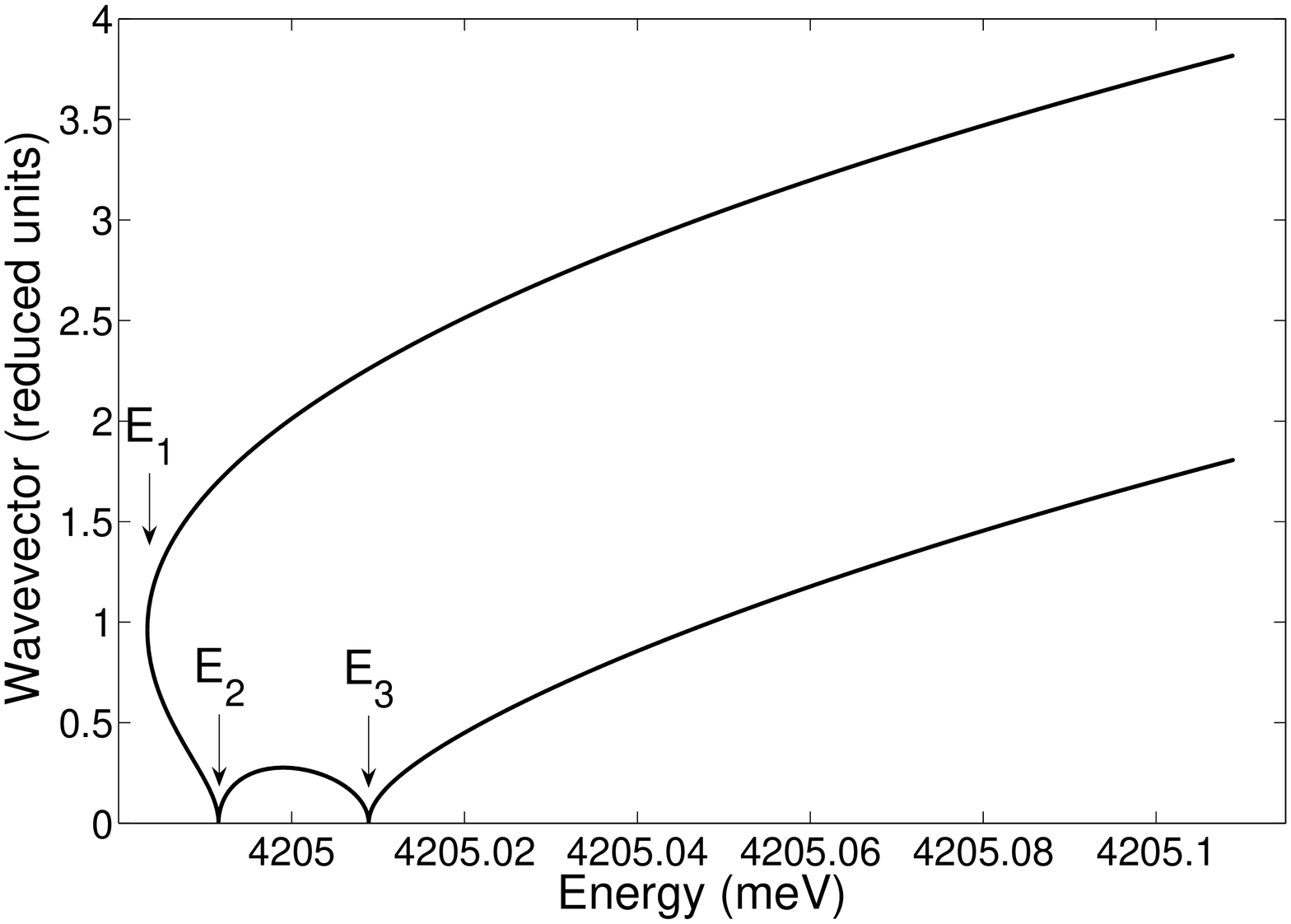,height=2.5in,width=3.7in,angle=-90}}
\end{figure}
\
\vskip .1in
\begin{figure}[h]
\centerline{\psfig{figure=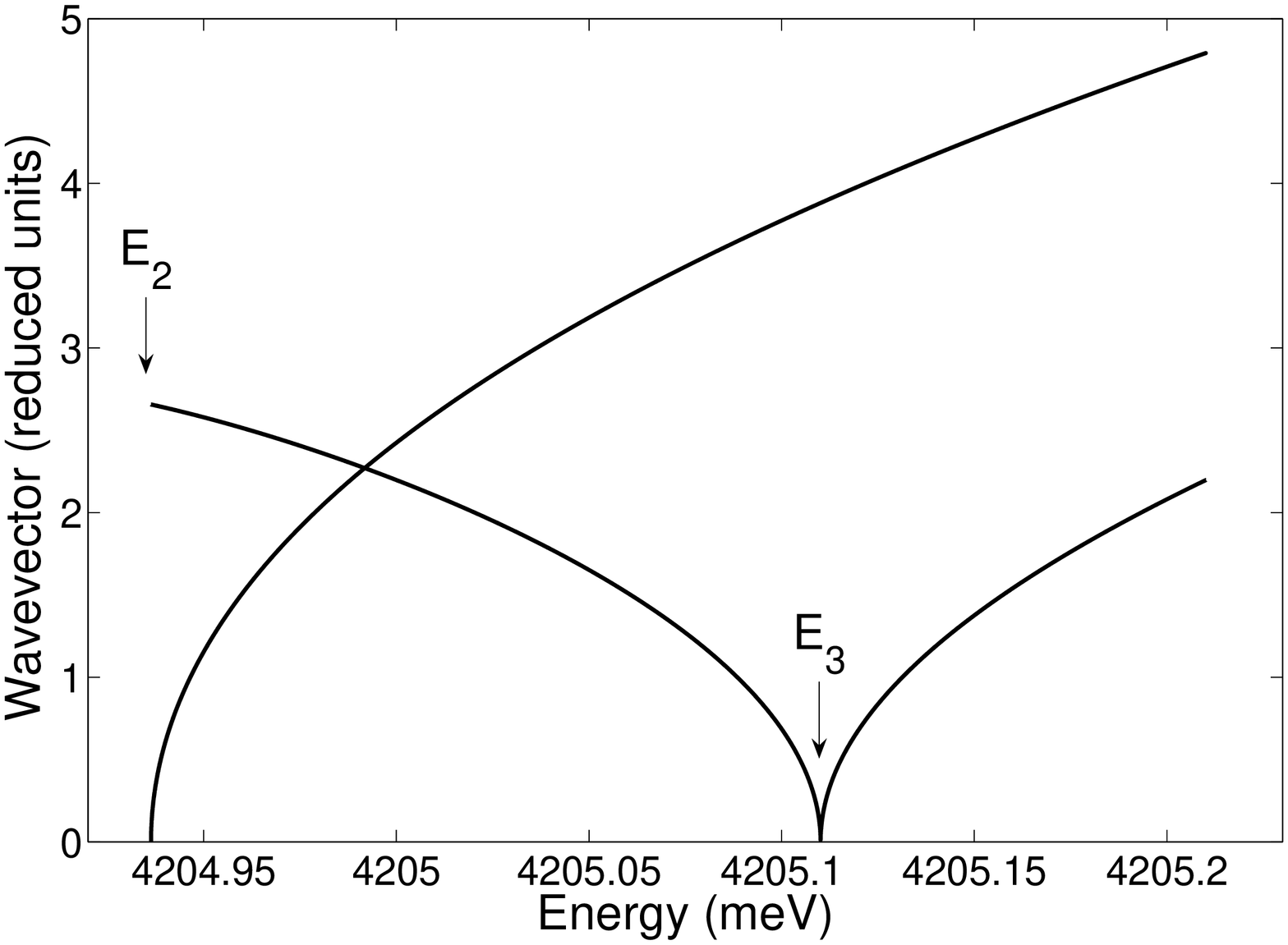,height=2.5in,width=3.7in,angle=-90}}
\vskip .5in
\begin{center}
Figure 4
\end{center}
\end{figure}

\newpage
\
\vskip .1in
\begin{figure}[h]
\centerline{\psfig{figure=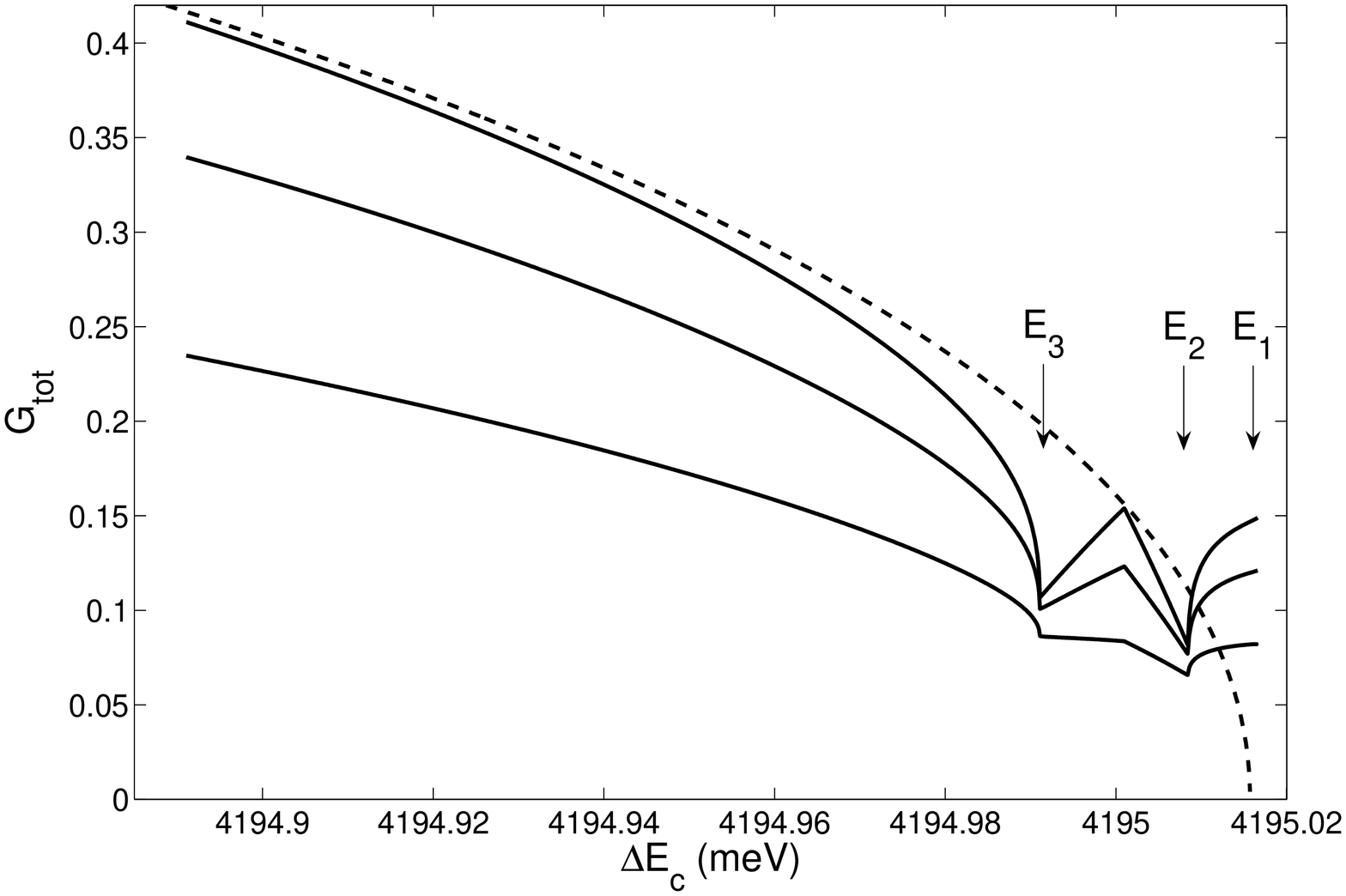,height=5in,width=5in,angle=-90}}
\vskip .5in
\begin{center}
Figure 5
\end{center}
\end{figure}

\newpage
\
\vskip .1in
\begin{figure}[h]
\centerline{\psfig{figure=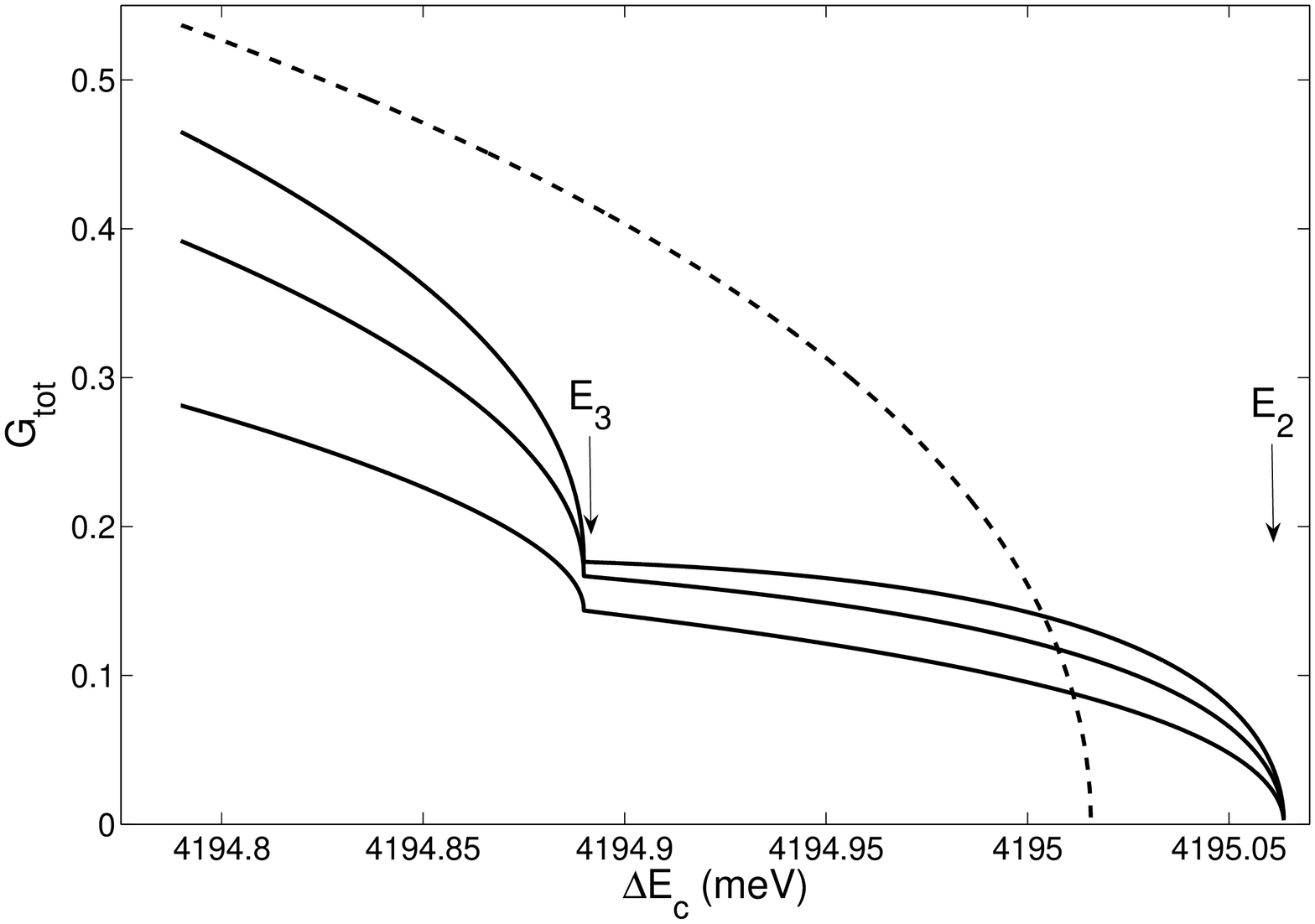,height=5in,width=5in,angle=-90}}
\vskip .5in
\begin{center}
Figure 6
\end{center}
\end{figure}

\newpage
\
\vskip .1in
\begin{figure}[h]
\centerline{\psfig{figure=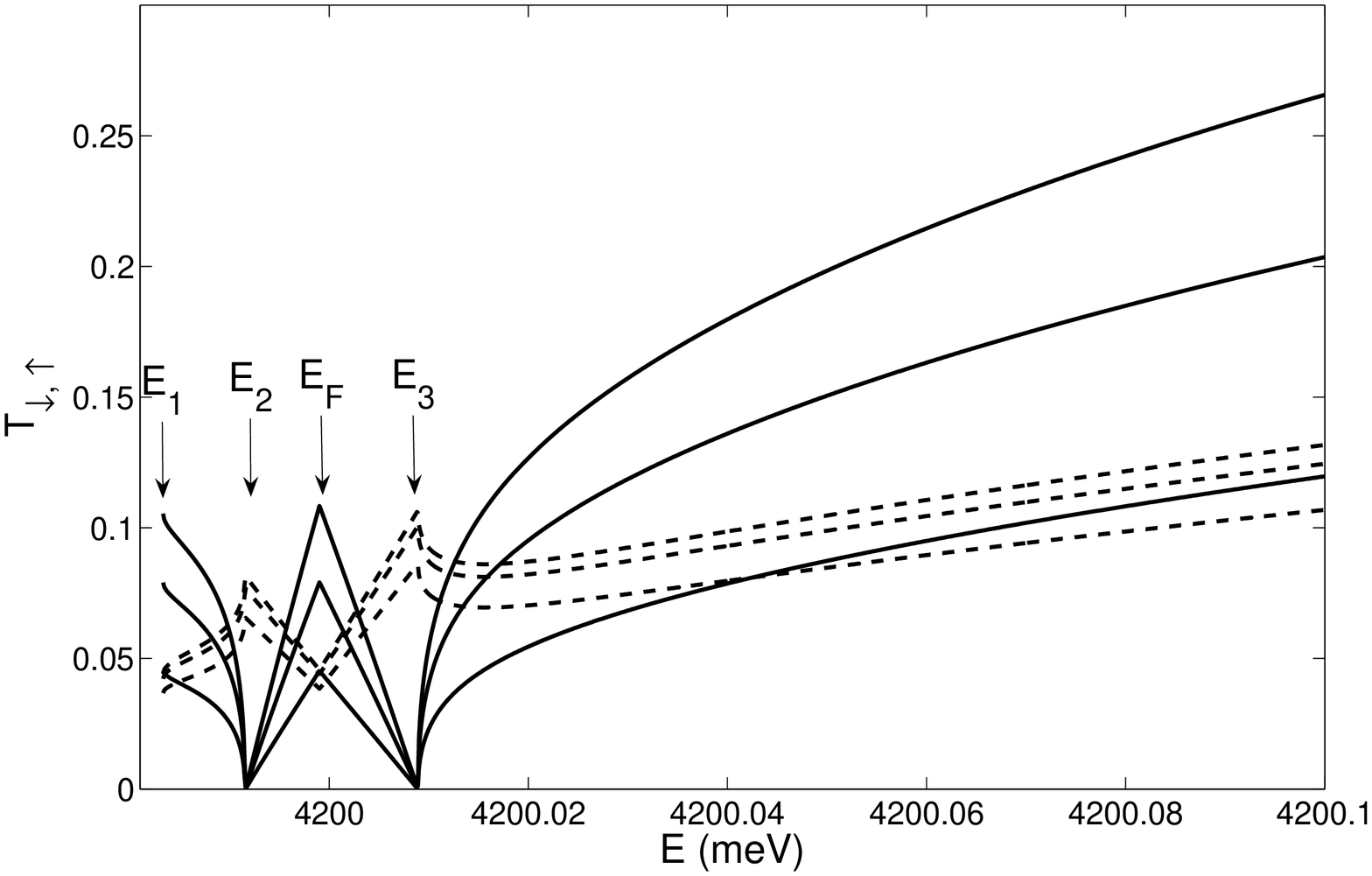,height=5in,width=5in,angle=-90}}
\vskip .5in
\begin{center}
Figure 7
\end{center}
\end{figure}

\newpage
\
\vskip .1in
\begin{figure}[h]
\centerline{\psfig{figure=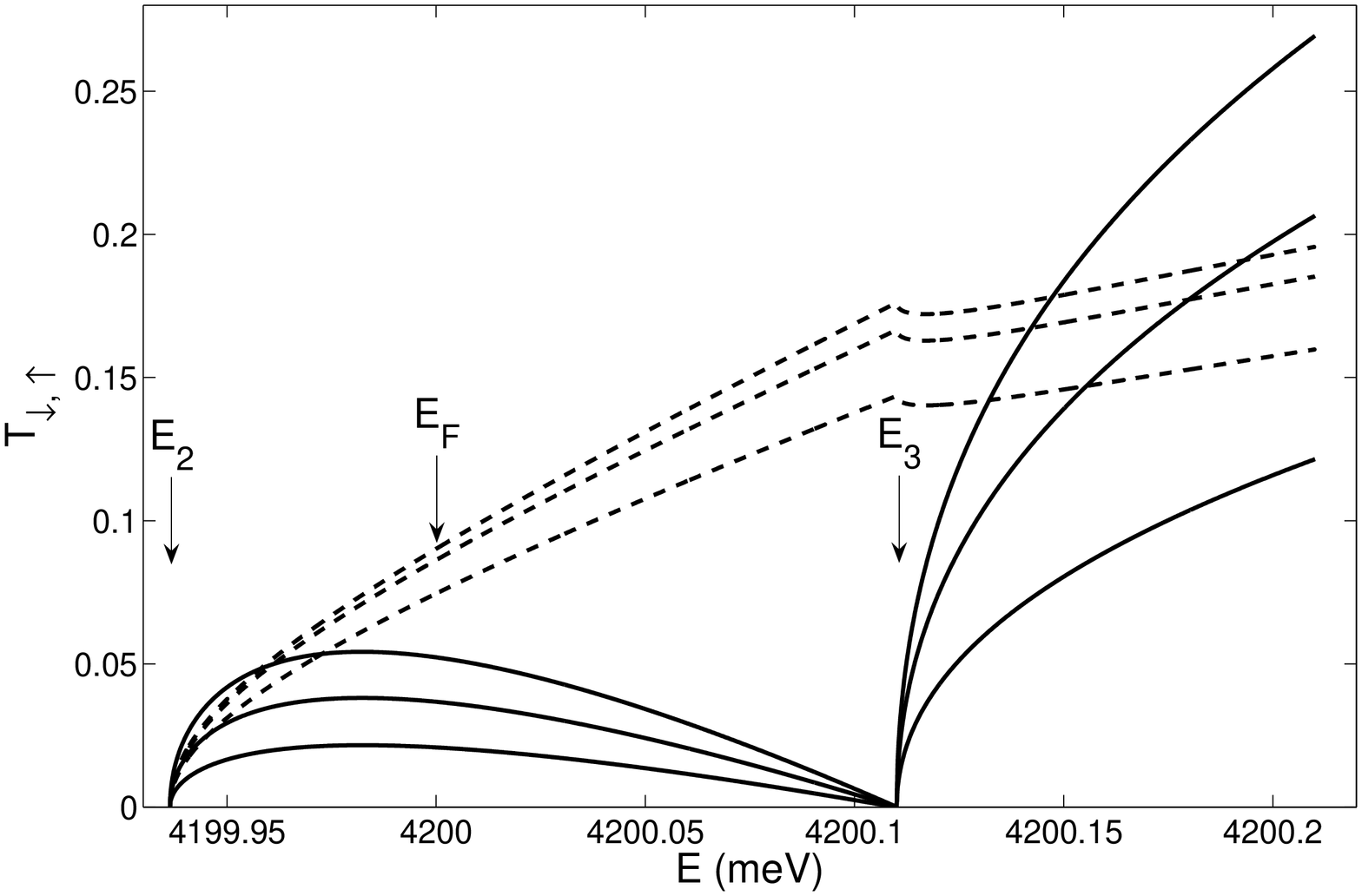,height=5in,width=5in,angle=-90}}
\vskip .5in
\begin{center}
Figure 8
\end{center}
\end{figure}

\newpage
\
\vskip 0.5in
\begin{figure}[h]
\centerline{\psfig{figure=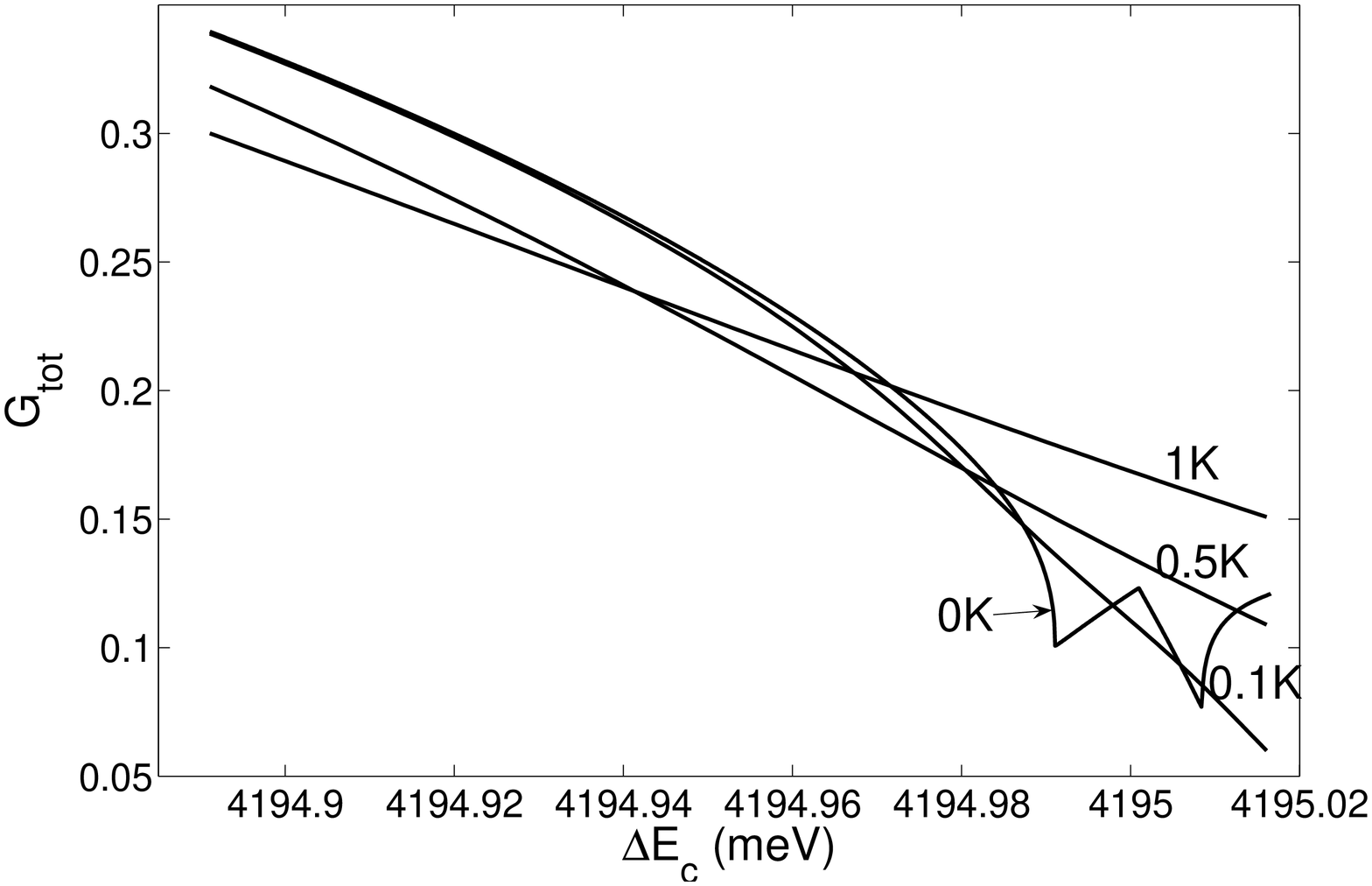,height=2.5in,width=3.7in,angle=-90}}
\end{figure}
\
\vskip .1in
\begin{figure}[h]
\centerline{\psfig{figure=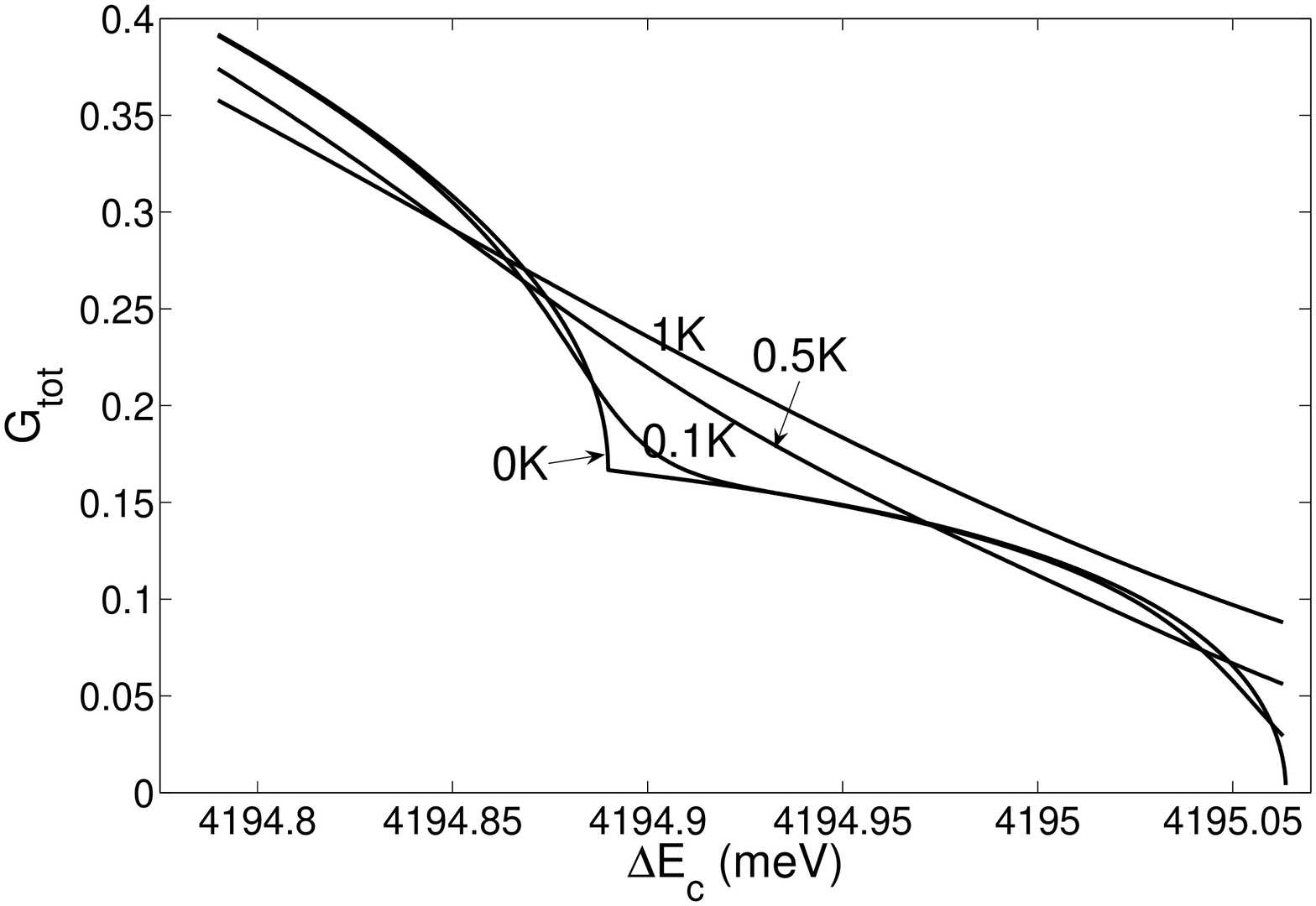,height=2.5in,width=3.7in,angle=-90}}
\vskip .5in
\begin{center}
Figure 9
\end{center}
\end{figure}

\newpage
\
\vskip .1in
\begin{figure}[h]
\centerline{\psfig{figure=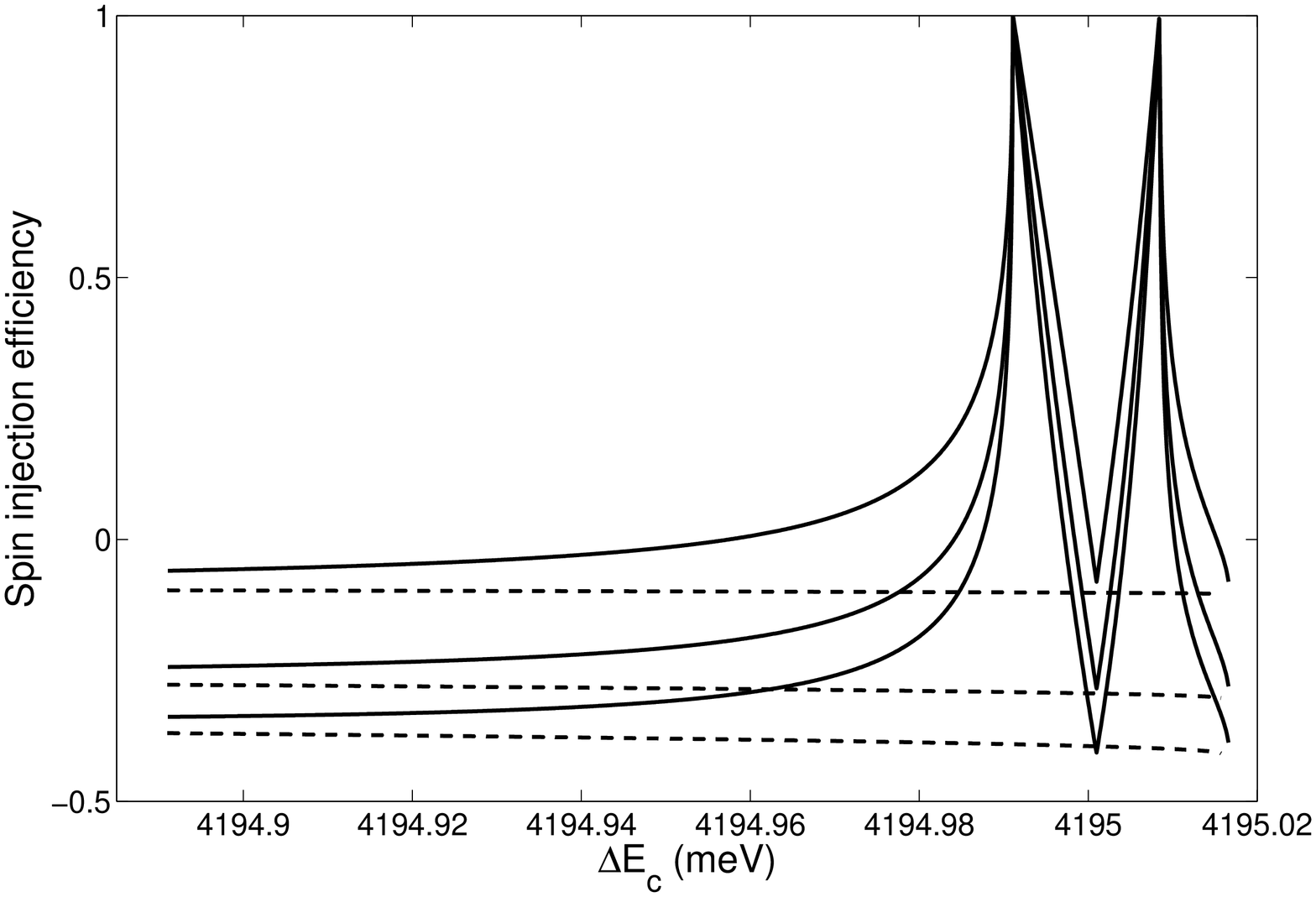,height=5in,width=5in,angle=-90}}
\vskip .5in
\begin{center}
Figure 10
\end{center}
\end{figure}

\newpage
\
\vskip .1in
\begin{figure}[h]
\centerline{\psfig{figure=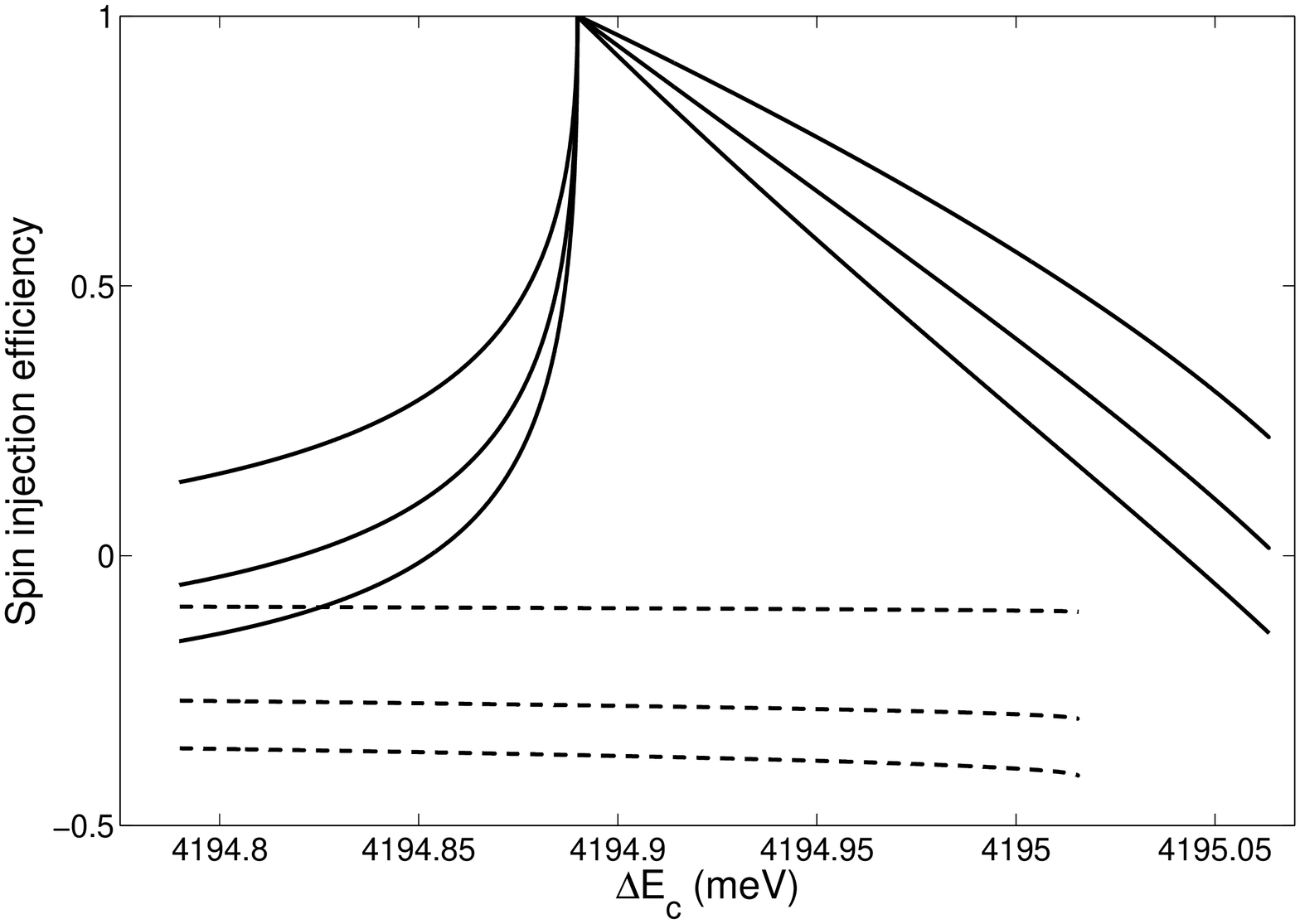,height=5in,width=5in,angle=-90}}
\vskip .5in
\begin{center}
Figure 11
\end{center}
\end{figure}

\newpage
\
\vskip .1in
\begin{figure}[h]
\centerline{\psfig{figure=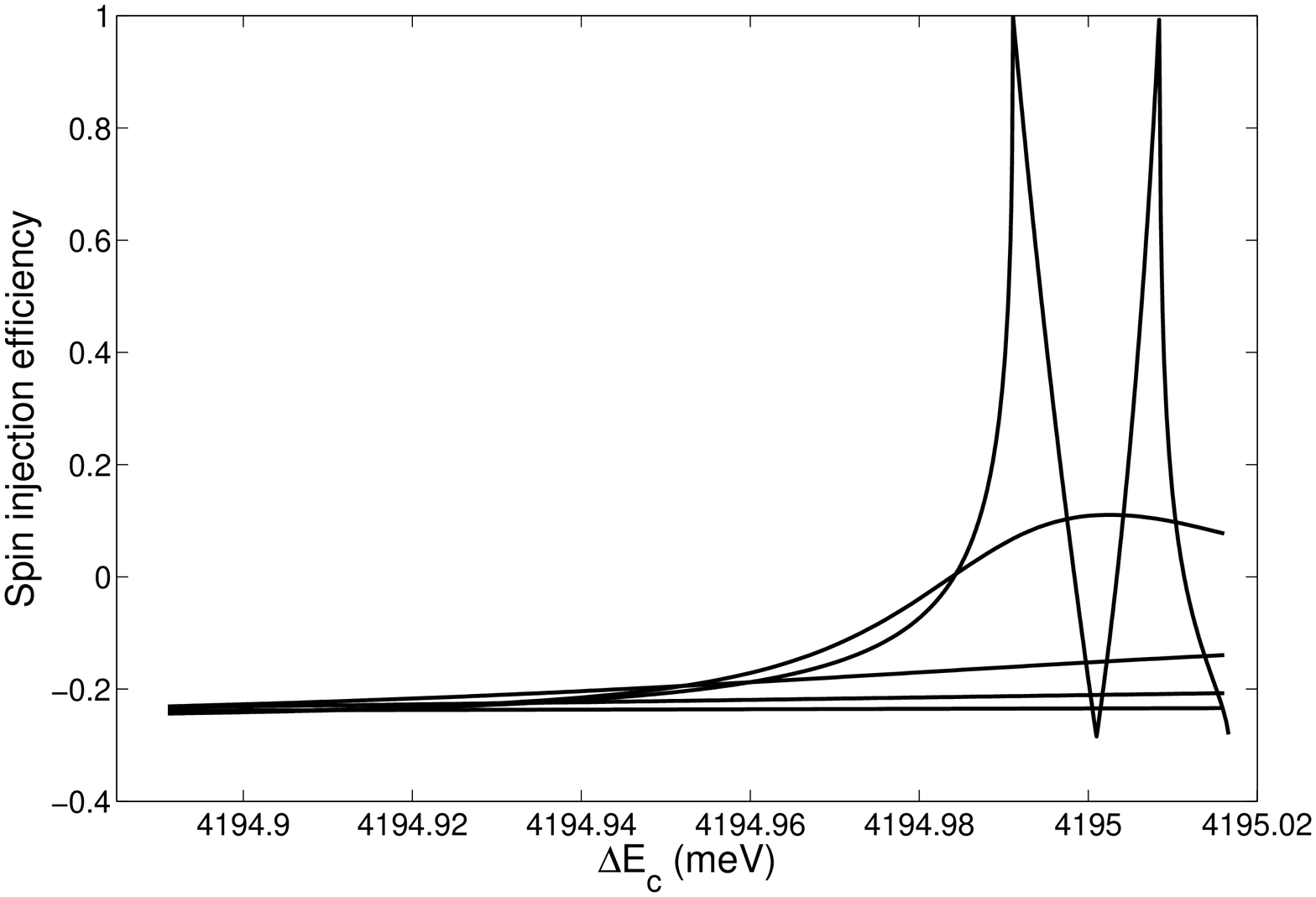,height=5in,width=5in,angle=-90}}
\vskip .5in
\begin{center}
Figure 12
\end{center}
\end{figure}

\newpage
\
\vskip .1in
\begin{figure}[h]
\centerline{\psfig{figure=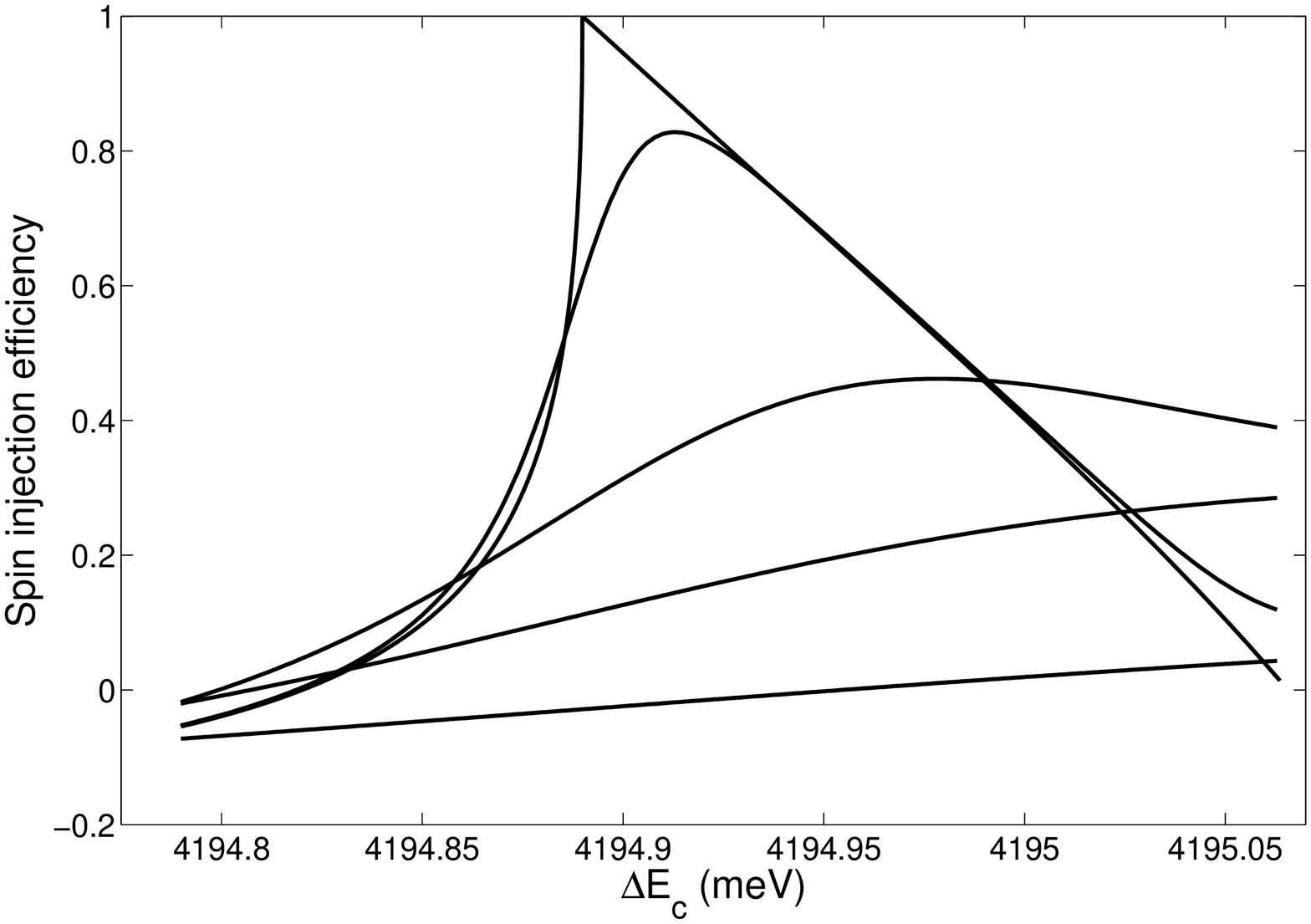,height=5in,width=5in,angle=-90}}
\vskip .5in
\begin{center}
Figure 13
\end{center}
\end{figure}

\newpage
\
\vskip .1in
\begin{figure}[h]
\centerline{\psfig{figure=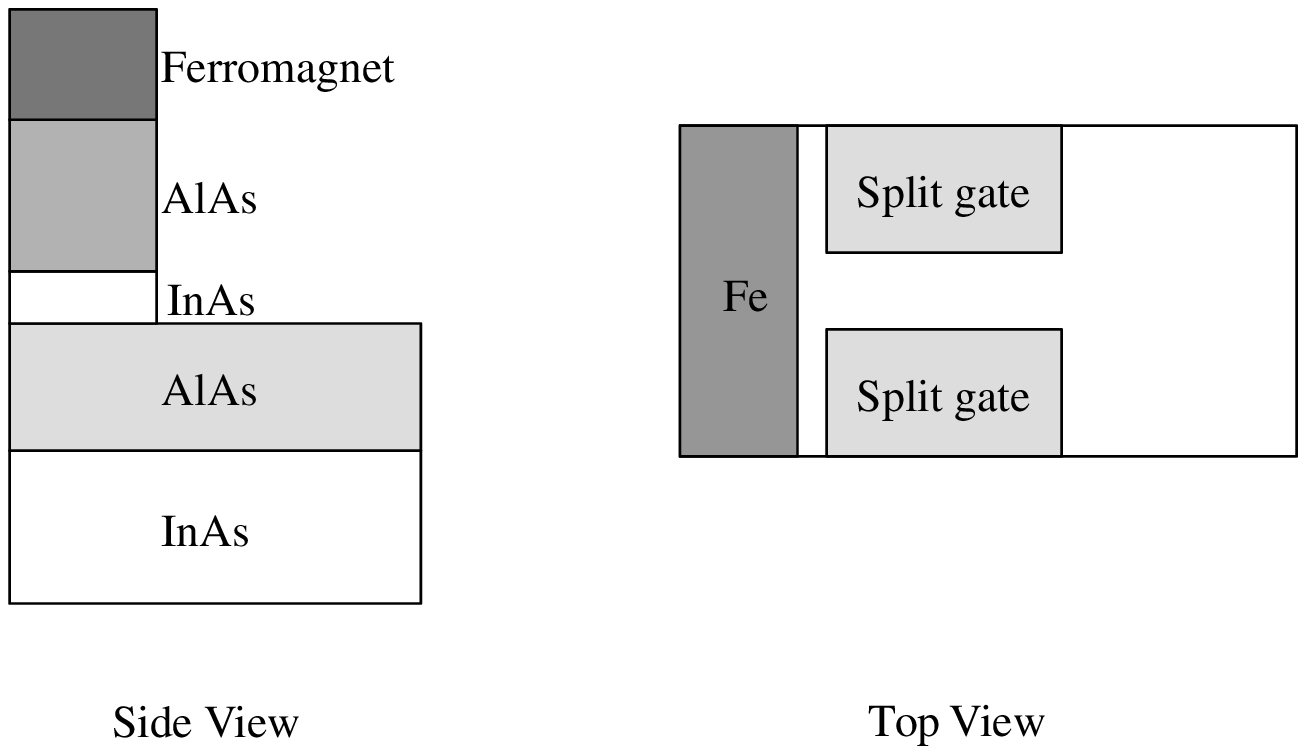,height=6.5in,width=6.5in}}
\vskip .2in
\begin{center}
Figure 14
\end{center}
\end{figure}


\begin{thebibliography}{10}
\bibitem{gstar}
For the most commonly studied III-V semiconductor materials for spintronics
applications (GaAs, InAs, and InSb), the $g^*$ values for bulk samples are negative and
equal to -0.44, -14.9, and -51.0, respectively.
\end{thebibliography}
\end{document}